\documentclass[%
reprint,
superscriptaddress,
amsmath,amssymb,
aps,
pra,
floatfix,showpacs
]{revtex4-1}
\usepackage{epsfig}
\usepackage{epstopdf}
\usepackage{graphicx}
\usepackage{amsfonts,amsmath,latexsym,amssymb}
\usepackage[usenames, dvipsnames]{color}
\usepackage{bbm}
\usepackage[hidelinks]{hyperref}
\usepackage{balance}

\newcommand*{\rom}[1]{\expandafter\@slowromancap\romannumeral #1@}
\newcommand{\RN}[1]{%
  \textup{\uppercase\expandafter{\romannumeral#1}}%
}
\def\be{\begin{equation}}
\def\ee{\end{equation}}
\def\bea{\begin{eqnarray}}
\def\eea{\end{eqnarray}}

\begin{document}

\def\dy227{Dy$_2$Ti$_2$O$_7$}
\def\ket#1{\vert #1 \rangle}
\def\bra#1{\langle #1 \vert}

\title{Slow relaxation and sensitivity to disorder in trapped lattice fermions after a quench}

\author{M. Schulz}

\affiliation{SUPA, School of Physics and Astronomy, University of St Andrews, North Haugh, St Andrews KY16\ 9SS, UK}
\affiliation{Max Planck Institut for the Physics of Complex Systems, Dresden, Germany}

\author{C. A. Hooley}

\affiliation{SUPA, School of Physics and Astronomy, University of St Andrews, North Haugh, St Andrews KY16\ 9SS, UK}

\author{R. Moessner}

\affiliation{Max Planck Institut for the Physics of Complex Systems, Dresden, Germany}

\date{7th December 2016}

\begin{abstract}
We consider a system of non-interacting fermions in one dimension subject to a single-particle potential consisting of (a) a strong optical lattice, (b) a harmonic trap, and (c) uncorrelated on-site disorder.  
After a quench, in which the center of the harmonic trap is displaced, we study the occupation function of the fermions and the time-evolution of experimental observables.  
Specifically, we present numerical and analytical results for the post-quench occupation function of the fermions, and analyse the time-evolution of the real-space density profile.  
Unsurprisingly for a non-interacting (and therefore integrable) system, the infinite-time limit of the density profile is non-thermal.  
However, due to Bragg-localization of the higher-energy single-particle states, the approach to even this non-thermal state is extremely slow.  
We quantify this statement, and  show that it implies a sensitivity to disorder parametrically stronger than that expected from Anderson localization.
\end{abstract}

\pacs{05.30.-d, 37.10.Jk, 67.85.Lm}

\maketitle

\section{Introduction}
\noindent
In the past twelve years or so there has been a significant increase in the level of theoretical activity on questions of thermalization, especially for isolated quantum systems \cite{Gornyi2005,Basko2006,Rigol2008,Polkovnikov2011,Huse2014a,Nandkishore2015,Eisert2015}.  
There are several reasons for this.  One is the growing availability of experimental realizations, for example in cold-atom systems \cite{Kinoshita2006,Billy2008,Roati2008,Schreiber2015,Kondov2015} and in the nuclear spins of solid-state dopants \cite{Childress2006}.  Another is the growing theoretical understanding of how quantum-mechanical systems approach thermal equilibrium.

The key concept in the classical statistical mechanics of isolated systems is ergodicity, which essentially depends on chaos. 
For quantum systems, it has been conjectured that a {\it single\/} typical many-body eigenstate of energy $E$ already matches the microcanonical ensemble in the expectation values it gives for local observables.  
This claim, related to Berry's conjecture \cite{Berry1977}, is referred to as the {\it eigenstate thermalization hypothesis\/} (ETH) \cite{Deutsch1991,Srednicki1994,Rigol2008}.

For this hypothesis to be true, nearby states in the many-body spectrum must have similar values of all local observables.  
However, one can easily think of examples in which this is not the case.  One class of these occurs in disordered systems, where the single-particle eigenfunctions are localized, and hence small changes in the total energy may lead to dramatic rearrangements of the spatial density profile.  
It has been discovered more recently that this idea extends to the case of interacting particles, where it goes by the name of {\it many-body localization\/} (MBL) \cite{Gornyi2005,Basko2006,Polkovnikov2011,
Huse2014a,Nandkishore2015,Eisert2015,Pal2010,Bardarson2012,Vosk2013,Chandran2014,Ponte2015,Agarwal2015,Imbrie2016}.

A second class occurs in integrable systems, where the number of conserved quantities is so large that neighboring states in the many-body spectrum are very likely to have different values of many of them, and thus disagree on many of their local observables.  
It is believed that this can be addressed by restricting the microcanonical ensemble to a distribution in which additional temperature-like parameters are introduced to constrain the values of these conserved quantities:\ the so-called {\it generalized Gibbs ensemble\/} (GGE) \cite{Rigol2007,Caneva2011,Cassidy2011,Caux2012,Caux2013,Pozsgay2014,Essler2015}.  
That said, it is not always clear how to properly enumerate the conserved quantities that should be included in such a modified ensemble.

The vast majority of work in this area considers systems with (continuum- or lattice-)translationally invariant Hamiltonians.  
However, the most popular experimental realizations, using cold atomic gases, generally involve a spatially inhomogeneous trapping potential.  
This suggests that it would be worth considering the influence of such a potential --- clean or disordered --- on the relaxation of a many-fermion system after a quench.  
This question is theoretically interesting because it concerns a quantum system relaxing under the influence of bulk but inhomogeneous forces.  
It is also of interest because of its direct relevance to experiment:\ indeed, reports of experiments exhibiting two or even all three of these ingredients (lattice, trap, disorder) may already be found in the literature \cite{Billy2008,Lye2005,HeckerDenschlag2002,Jordens2008,Schulte2005,White2009,Choi2016}.

\begin{figure}
\centerline{\includegraphics[width=1.0\columnwidth]{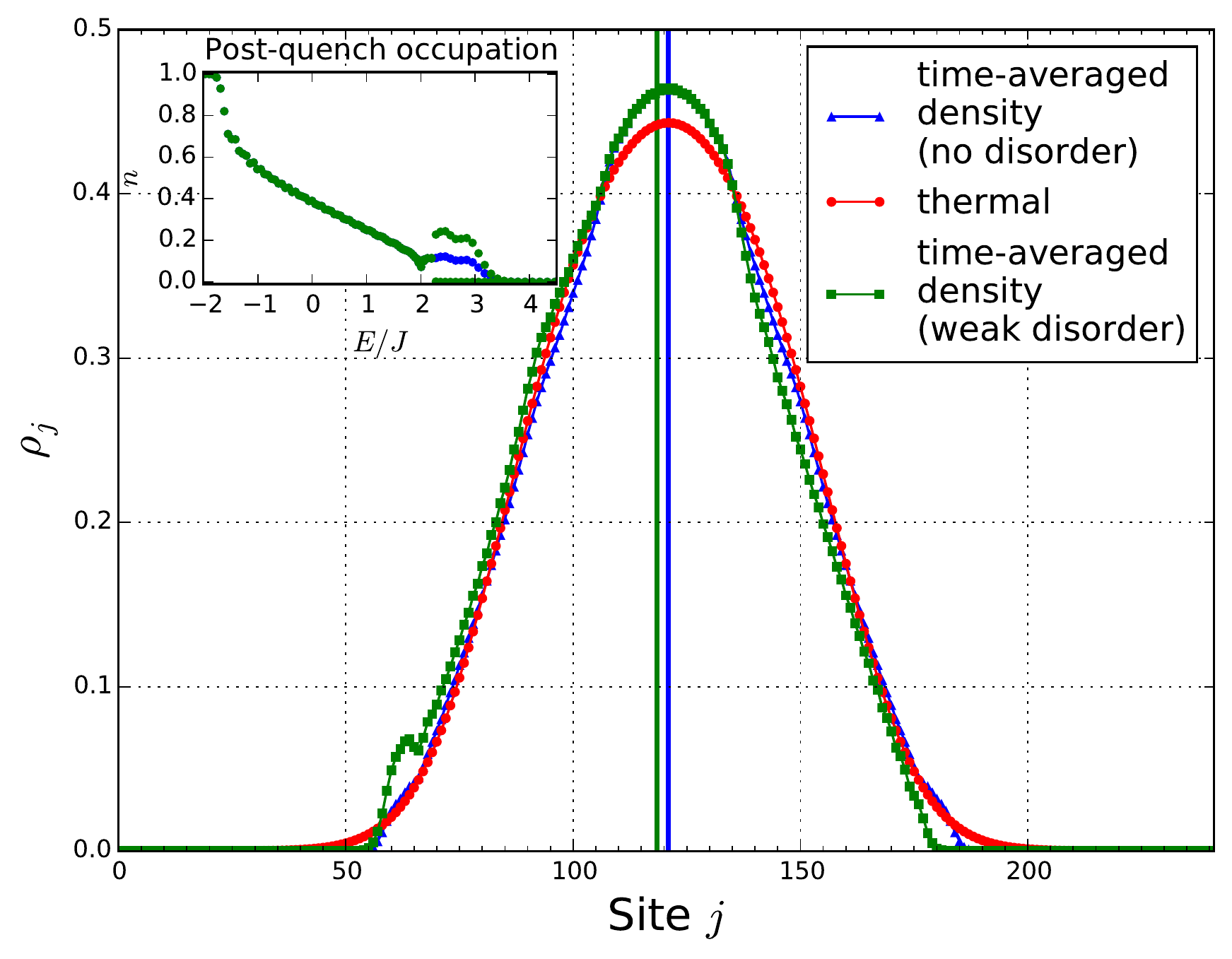}}
\caption{Main panel:\ We compare the real-space time-averaged density profile of the fermions without disorder (blue curve with triangles) and in the case of very weak disorder (green curve with squares) to the equilibrium prediction for a system with the same number of particles and the same total energy (red curve with circles).
Note that even the center of mass of the weakly disordered time-averaged profile (solid green line) does not coincide with the equilibrium or clean case prediction (solid blue line).
Inset:\ The occupation function of the post-quench states ordered by their energy in the weakly disordered (green) and clean (blue) cases.
Note the occupation of states with energies $E > 2J$:\ these states are Bragg-localized.
Parameters:\ number of lattice sites $L = 241$; trap spring constant $\kappa = 0.0025$; hopping integral $J = 1$; chemical potential $\mu =0$; pre-quench trap center $j_{0} = 96$; post-quench trap center $j_{1} = 121$; disorder strength $W =\{0,10^{-5}\}$.}
\label{fig:dens_inf_dis}
\end{figure}

In this work, we consider a global quantum quench applied to a one-dimensional system of spinless, non-interacting fermions in a potential consisting of a strong optical lattice, a harmonic trap, and sometimes also uncorrelated site disorder.
The quench protocol consists of letting the system equilibrate, and then, at the moment of the quench, suddenly displacing the center of the harmonic trapping potential from its initial position by $\Delta j$ lattice sites.
Such quenches were first studied experimentally over a decade ago \cite{Pezze2004,Ott2004,Modugno2003}.
We investigate the representation of the pre-quench state in the post-quench eigenbasis, which is the initial condition for all subsequent time-evolution.  
We also analyze how that time-evolution affects the values of observables such as the moments of the fermions' spatial density profile.

Since our fermions are non-interacting, the population of each post-quench single-particle eigenstate is a constant of motion, and the system is trivially integrable.  
Nonetheless, as we change the trap-jump distance $\Delta j$ and the strength of the disorder $W$ we observe considerable variation in the timescales on which different observables relax to their time-averaged values, and in the extent to which those time-averaged values agree with equilibrium predictions.
A precise definition of the time-averaged density is given in Section~\ref{sec:time}.

For example, for large enough trap jumps, even when the disorder is extremely weak, we find that the violation of parity present in the initial conditions is preserved in the infinite-time ($t \to \infty$) density profile.  
This represents a dramatic failure to match the form predicted by equilibrium statistical mechanics (see Fig.~\ref{fig:dens_inf_dis}).
This is not due to Anderson localization.  
Rather, it is associated with the extreme disorder-sensitivity of the
Bragg-localized states in the upper part of the single-particle
spectrum.
For clarity, we note in passing that Bragg localization and Anderson localization are conceptually quite distinct.  While the latter is defined as the absence of diffusion in the presence of randomness, the former occurs in a setting here where such a definition is not natural, because the unbounded trap potential in any case eventually prevents diffusion.  A natural description of Bragg localization is rather that there are high-lying eigenstates that are exponentially localized on a shorter length-scale than the classically allowed region set by the trap.  While we use uncorrelated on-site disorder for simplicity, any term in the Hamiltonian that breaks the parity symmetry, e.g.\ an Aubry-Andr{\'e} potential or even a non-integer post-quench trap position $j_1$, would yield analogous effects.

\begin{figure}
\centerline{\includegraphics[width=\columnwidth]{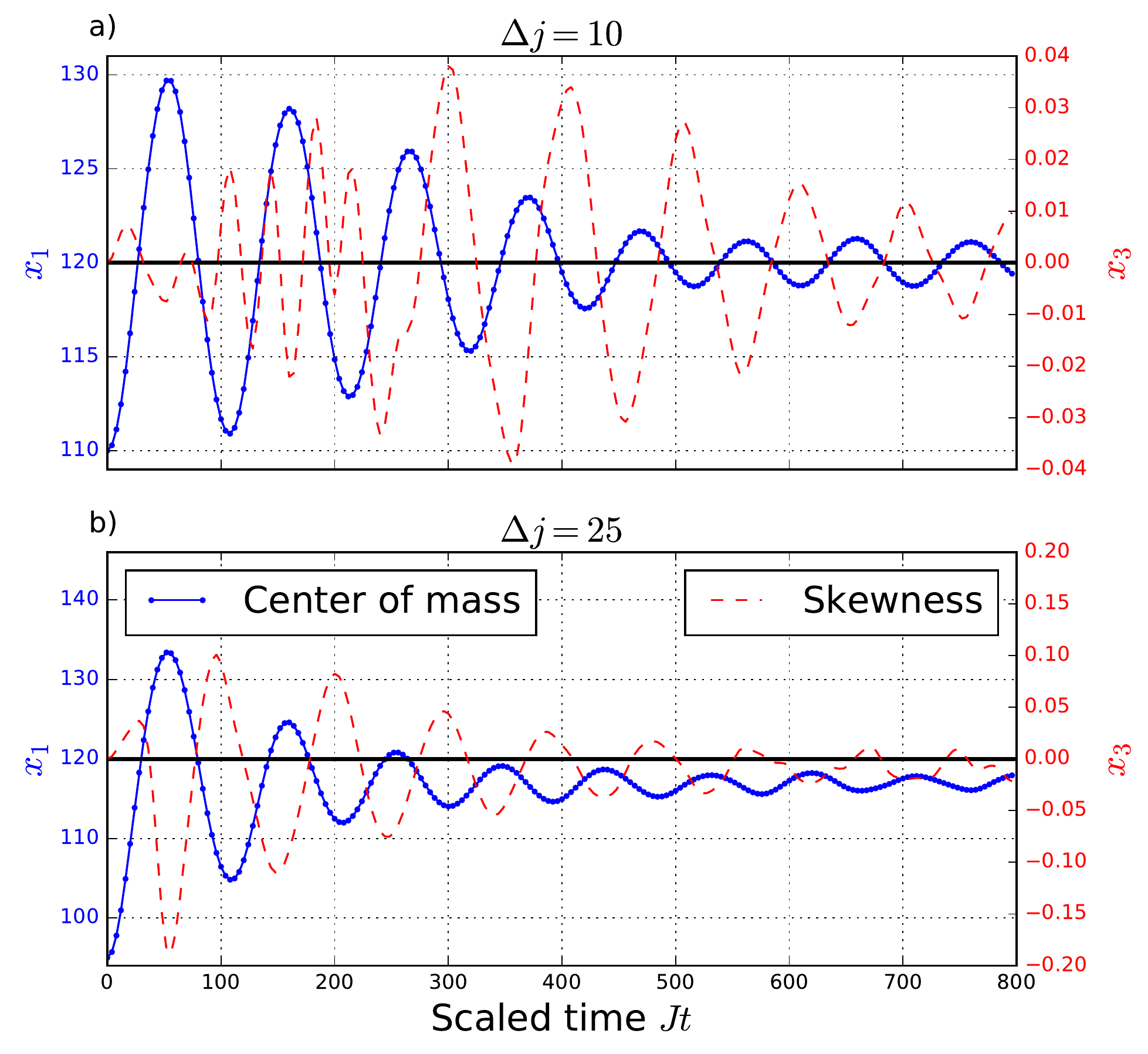}}
\caption{(a) Center of mass $x_1$ (blue curve with circles; left-hand scale) and skewness $x_3$ (red dashed curve; right-hand scale) as functions of time, for a quench with trap-jump size $\Delta j = 10$.  The center of mass oscillates around the post-quench trap center (solid black line).  (b) The same, but for a larger trap-jump size $\Delta j = 25$.  Again, the post-quench trap center is indicated by the solid black line; but now, even though there is no disorder, on observable time-scales the center of mass instead oscillates around a different point, between pre- and post-quench trap centers.
Parameters:\ number of lattice sites $L = 241$; trap spring constant $\kappa = 0.0025$; hopping integral $J = 1$; chemical potential $\mu = 0$; post-quench trap center $j_{1} = 120$; disorder strength $W = 0$.}
\label{fig:CoM_Skew}
\end{figure}

In the complete absence of disorder, parity is eventually restored by the dephasing of these Bragg-localized states, but the timescale on which this occurs is extremely long.  
Thus on experimentally relevant timescales the clean case is not actually materially different from the disordered one.  
In both, for example, the center of mass oscillates not about the new center of the trap, but about a point between the pre- and post-quench trap centers (see Fig.~\ref{fig:CoM_Skew}).
The question whether the center of mass reaches the new trap center, and in particular the role of Bragg localized states~\cite{Rigol2004} and the existence of parity doublets~\cite{Ruuska2004}, was already raised following the original experiment~\cite{Modugno2003}.

The plan of the remainder of this paper is as follows. In Sec.\ \ref{sec:model}, we introduce the model and discuss the quench protocol. 
In Sec.\ \ref{sec:distribution}, we analyze the representation of the pre-quench state in the post-quench basis --- the initial condition for the post-quench time-evolution --- for a range of trap-jump sizes $\Delta j$.
Sec.\ \ref{sec:time} provides an analysis of the time-evolution of the moments of the density and investigates the short and long time properties of the density itself. We also include the influence of disorder on the dynamics, elucidating the competition between the two forms of localization in the system. 
We conclude with Sec.\ \ref{sec:conclusion}, in which we briefly summarize our results, and also discuss possible future developments, especially the introduction of atom-atom interactions and the associated questions of many-body localization.

\section{Model and quench protocol}
\label{sec:model}
\noindent
We consider spinless fermions moving in one dimension on a lattice of $L$ sites with open boundary conditions.
The Hamiltonian reads
\begin{multline}
\label{equ:hamiltonian}
\hat{H}_{i} = -J\sum_{j = 1}^{L-1}\left(c^{\dagger}_{j}c_{j+1}+ c^{\dagger}_{j+1}c_{j}\right)+ \\ \sum_{j=1}^{L}\left[\frac{1}{2}\kappa a^2 \left(j-j_{0}\right)^2+\epsilon_{j}\right]c^{\dagger}_{j}c_{j}.
\end{multline}
Here the operator $c^{\dagger}_{j}$ creates a fermion on site $j$, and $J$ is the hopping matrix element between neighboring sites.
The on-site energy consists of a harmonic trapping potential of spring constant $\kappa$ centered at $j_{0}$ plus additional uncorrelated on-site disorder taken from a uniform box distribution: $\epsilon_{j} \in \left[-W,W\right]$.
For convenience we shall henceforth set both $\hbar$ and the lattice constant $a$ to unity.
We find the single-particle eigenstates $\{ \alpha_{k}\}$ of $\hat{H}_{i}$ and populate the lowest $N$ of them to obtain the initial ground state of the $N$-fermion problem.
Alternatively we can choose a chemical potential $\mu$ and populate all single-particle eigenstates for which the eigenenergy $E_{k}^{(i)}$ is smaller than $\mu$.

In this paper, we study the non-equilibrium dynamics of this model that arise from a particular spatially inhomogeneous global quench.
At time $t=0$ the center of the harmonic trapping potential is displaced from site $j_0$ to site $j_1$, while the disorder potential is left unchanged.
Thus the post-quench Hamiltonian $\hat{H}$ is exactly the same as \eqref{equ:hamiltonian} but with $j_{0} \rightarrow j_{1}$.
This Hamiltonian has a set of single-particle eigenstates $\{ \beta_{k}\}$ with eigenenergies $E_{k}$.  We define the `jump size' $\Delta j$ as $\vert j_1 - j_0 \vert$.

The subsequent time-evolution of the many-body state of the system can be understood as a dephasing of the contributions of the individual post-quench eigenstates, due to their different eigenenergies.
The pre-quench state, represented in the post-quench basis, serves as the initial condition for this time-evolution.  In the coming sections, we study further the nature of this initial condition, and of the subsequent time-evolution of physical observables such as the center-of-mass of the atom cloud.

\section{The post-quench occupation function}
\label{sec:distribution}

\begin{figure}
\centerline{\includegraphics[width=\columnwidth]{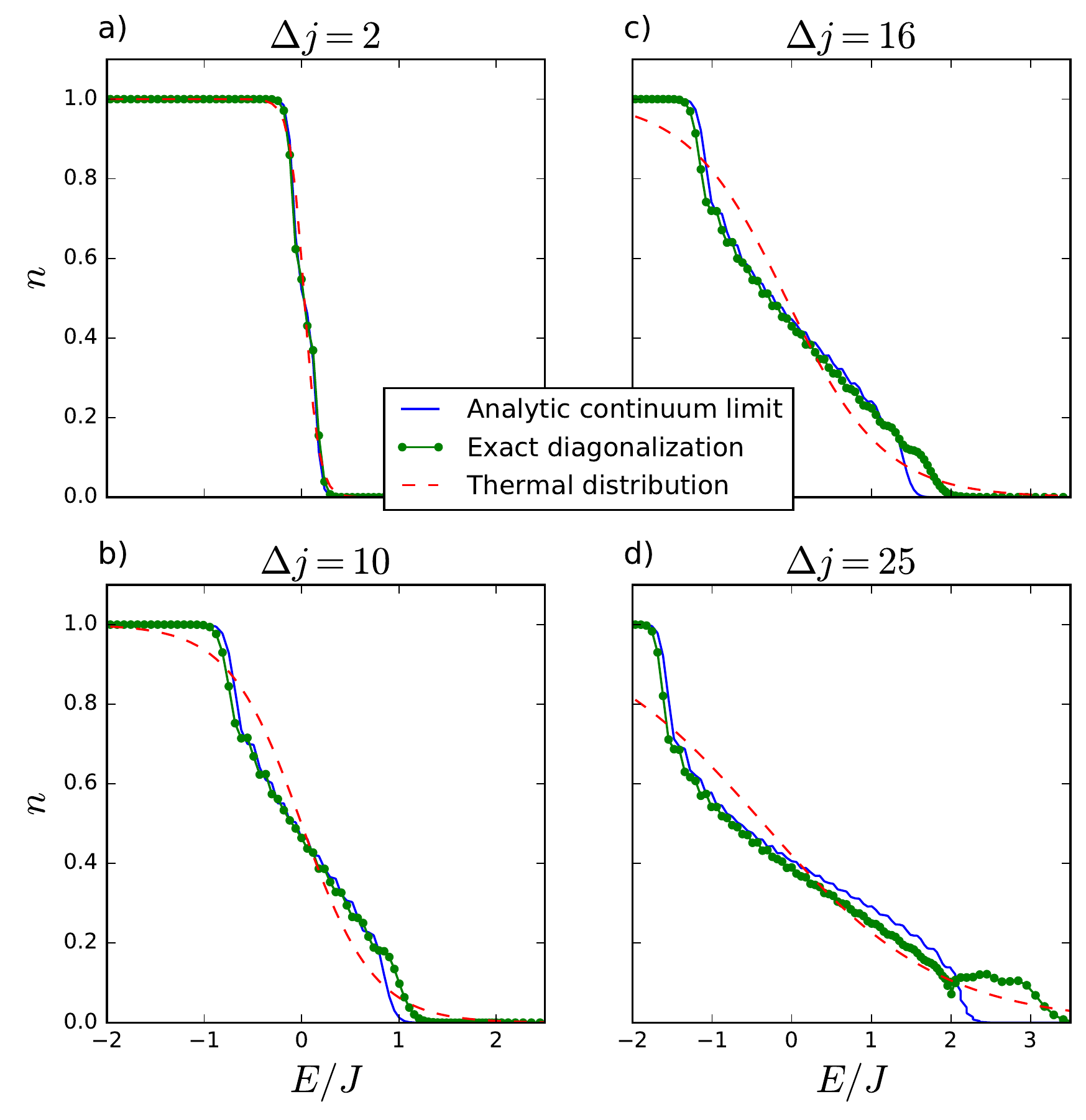}}
\caption{The post-quench occupation function $n(E)$ (green; circles) for four different jump sizes $\Delta j$.  For comparison we also plot the continuum result (blue; solid line) and the result for a thermal state with the same total energy and particle number (red; dashed line).
For small $\Delta j$ (panels (a) and (b)), the continuum approximation is a good one.
As soon as we start populating states above $E = 2J$, i.e. Bragg localized states, the continuum approximation fails (panels (c) and (d)).
Parameters:\ number of lattice sites $L = 241$; trap spring constant $\kappa = 0.0025$; hopping integral $J = 1$; chemical potential $\mu =0$; post-quench trap center $j_{1} = 121$; disorder strength $W =0$.}
\label{fig:distr_hop_multjump}
\end{figure}

\noindent
In order to study the time-evolution of the system for times $t>0$, we need to know the state at time $t=0$, i.e.\ we need to represent the pre-quench state in the post-quench basis.  This will consist of a superposition of many different Slater determinants, each corresponding to a different assignment of the $N$ fermions to the $L$ post-quench single-particle eigenstates.  A simple function that captures its essence, however, is the expectation value of the occupation of each post-quench single-particle eigenstate, $n_{k}^{\left(\beta\right)}$.  Here $k=1,2,\ldots,L$ labels the post-quench single-particle eigenfunctions.

Since the fermions are non-interacting, each $n_{k}^{\left(\beta\right)}$ is a constant of the motion.  This trivially prevents the system from thermalizing; nonetheless, particular observables --- e.g.\ the center of mass of the atom cloud --- may still relax to their thermal equilibrium values.

In order to determine $n_{k}^{\left(\beta\right)}$ we express the pre-quench ground state $\vert \psi_{0}^{(N)} \rangle$ for $N$ particles as:
\begin{equation}
\label{equ:prequenchgs}
\vert \psi_{0}^{(N)} \rangle = \alpha_{N}^{\dagger} \alpha_{N-1}^{\dagger} \hdots \alpha_{2}^{\dagger} \alpha_{1}^{\dagger}\vert 0 \rangle,
\end{equation}
where the operator $\alpha^\dagger_k$ creates a fermion in pre-quench single-particle eigenstate $\alpha_k$, and $\ket{0}$ is the fermionic vacuum.
The post-quench occupation function is then defined as:
\begin{equation}
\label{equ:distrfunction}
n_{k}^{\left(\beta\right)} \equiv \langle \psi_{0}^{(N)} \vert\beta^{\dagger}_{k}\beta_{k}\vert \psi_{0}^{(N)} \rangle = \sum_{q=1}^{N}\left\vert O_{qk}\left(\Delta j\right)\right\vert^2,
\end{equation}
where $\beta^\dagger_k$ creates a fermion in post-quench single-particle eigenstate $\beta_k$, and the overlap matrix $O_{qk}(\Delta j)$ is defined as
\begin{equation}
\label{equ:overlap}
O_{qk}(\Delta j) \equiv \langle \alpha_{q} \vert \beta_{k} \rangle.
\end{equation}
Since there is a one-to-one mapping between the eigenstate quantum numbers $k$ and the eigenenergies $E_k$, we may equivalently represent the occupation function as $n(E)$, which we sample at the points $E=E_k$.

In Fig.~\ref{fig:distr_hop_multjump} we plot this post-quench occupation function for four different jump sizes.  For comparison, we show also the occupation function calculated in the continuum, and the thermal occupation function for the same total energy and particle number.

The post-quench occupation function exhibits a remarkable amount of structure.
Unlike the thermal distribution, it has a very steep slope when departing from zero and unity. For a wide range of small jump-sizes it also shows an almost linear structure around the Fermi energy $E_{F}=0$ that has a plateau-like substructure.

The continuum approximation works better for small trap jumps than for larger ones.  The reason is that small jumps mainly occupy low-lying single-particle eigenstates of the post-quench Hamiltonian.  These resemble the eigenstates of a continuum harmonic oscillator \cite{Hooley2004}.  Thus the `athermal' structure of the occupation function in these cases arises from the harmonic-oscillator nature of the eigenstates, rather than from the influence of the lattice.

For larger jump sizes, where the quench populates the higher-lying single-particle eigenstates, the continuum approximation becomes worse.
In particular, for energies $E \geqslant 2J$ the true occupation function and the continuum approximation to it disagree sharply.
This is because the single-particle eigenstates with energies $E \geqslant 2J$ are (to use the terminology of \cite{Hooley2004}) Bragg-localized:\ instead of extending between the two classical turning points, they go only as far as the atom can propagate before being Bragg-reflected from the optical lattice.

In the remainder of this section we expand on these observations, providing an exact-diagonalization study of the single-particle eigenstates, a detailed explanation of Bragg localization, and a derivation of the continuum approximation to the occupation function.  We shall focus on the dependence of the occupation function on the jump size $\Delta j$; the influence of the hopping strength $J$ and the disorder strength $W$ is discussed in Appendices \ref{app:occ_hop_dis} and \ref{app:hopping-and-disorder}.

\subsection{Post-quench single-particle eigenstates}
\label{sec:ED_eigenstates}

\begin{figure}
\centerline{\includegraphics[width=\columnwidth]{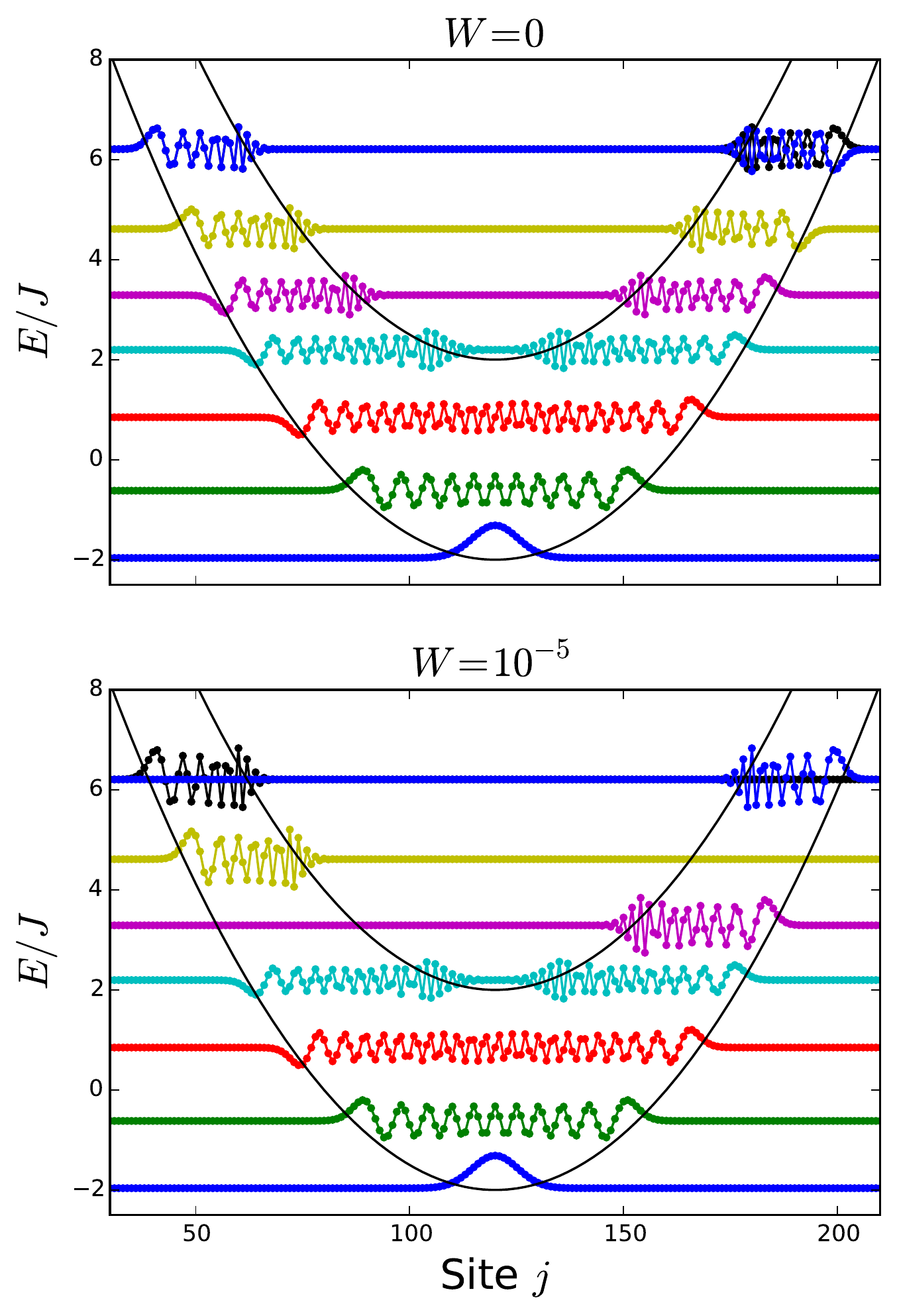}}
\caption{Selected post-quench single-particle eigenfunctions, determined by exact diagonalization.
Each eigenfunction is offset vertically by its eigenenergy.
The highest-lying eigenfunction is shown together with its almost degenerate partner.
The outer parabola shows the classical turning points as a function of energy while the inner parabola shows the Bragg turning points.
Upper panel:\ clean case ($W=0$).  Lower panel:\ very weak disorder ($W=10^{-5}$).
Parameters:\ number of lattice sites $L = 241$; trap spring constant $\kappa = 0.0025$; hopping integral $J = 1$; post-quench trap center $j_{1} = 121$.}
\label{fig:ED_eigenstates}
\end{figure}
\noindent
Fig.~\ref{fig:ED_eigenstates} shows some of the post-quench single-particle eigenstates, obtained by numerical exact diagonalization \footnote{In the cases where
the exponentially small energy splitting between the symmetric and antisymmetric eigenstates of the problem is beyond the resolution of
our numerical solver, we have `manually' taken linear combinations of the results to produce the correct symmetric and antisymmetric eigenstates}.
The nature of these eigenstates was first discussed in \cite{Rigol2004,Hooley2004}; here we briefly summarize their properties.

In the clean case, and in the absence of the harmonic trap, i.e.\ when $\kappa = W = 0$, our model is just a tight-binding model with a band dispersion $E(k) = -2J \cos k$, where $k$ is the wave number. In this limit the density of states is only non-zero for $\vert E \vert \leqslant 2J$, the region which we call the band. We will use the terminology of the band, especially `top' and `bottom' to refer to $E = \pm 2J$ respectively, even when $\kappa \neq 0$.

Adding a harmonic trap, i.e.\ setting $\kappa \ne 0$, imposes a finite spatial extent on the eigenfunctions.  As discussed in \cite{Hooley2004}, this may be determined semiclassically by considering the orbit of a particle whose total energy is given by $E = -2J \cos k + \kappa x^2/2$.
For $E<2J$, the orbit has only the conventional classical turning points, where $k=0$:
\begin{equation}
j = j_1 \pm j_c, \quad j_c = \sqrt{\frac{2E+4J}{\kappa}}.
\end{equation}
By contrast, for energies $E \geqslant 2J$ the orbit acquires in addition two Bragg turning points, where $k=\pm \pi$:
\begin{equation}
j = j_1 \pm j_b, \quad j_b = \sqrt{\frac{2E-4J}{\kappa}}. \label{jbe}
\end{equation}
Bragg reflection exponentially suppresses the wave function in the region between the two Bragg turning points. We call this region `Bragg-forbidden' and the states that exhibit such suppression `Bragg-localized'.  As can be seen in Fig.~\ref{fig:ED_eigenstates}, these turning points provide a good description of the spatial extent of the numerically determined eigenfunctions.

How are these eigenstates affected by the addition of disorder?  In the clean ($W=0$) case, the Hamiltonian $H$ is symmetric under a reflection about the trap center $j_1$.  Hence each eigenstate is either odd or even under such a reflection.  For energies $E$ well above $2J$, i.e.\ well into the Bragg-localized regime, each even eigenstate has an odd partner with almost the same energy.  These may be thought of as bonding and anti-bonding combinations of a left Bragg-localized and a right Bragg-localized state.  In the $E \to \infty$ limit, the energy splitting between the bonding and anti-bonding states tends to zero, and the left- and right-localized states become exact eigenstates of the problem.  But for any finite eigenenergy they are hybridized by a non-zero tunnelling matrix element $T$:
\begin{equation}
\label{equ:matelT}
T\approx e^{-j_{b}(E)/\xi(E)},
\end{equation}
where the decay length $\xi(E)$ is given approximately by
\begin{equation}
\xi(E) = -\frac{1}{\text{ln}\left(\frac{E}{2J}-\sqrt{\left(\frac{E}{2J}\right)^2-1}\right)}. \label{xie}
\end{equation}
(For details of the derivation of (\ref{equ:matelT}) and (\ref{xie}), see Appendix \ref{app:MatrixT}.)

However, the introduction of very weak disorder, $W \sim T$, is sufficient to suppress this hybridization, thus making the left- and right-localized states the true eigenstates of the problem.  This phenomenon is illustrated in Fig.~\ref{fig:ED_eigenstates}.  The two highest-energy eigenfunctions in the upper panel are the hybridization-split bonding and anti-bonding states; the two highest-energy eigenfunctions in the lower panel are the left- and right-localized states.  The hybridization between them has been suppressed even though the disorder strength is more than five orders of magnitude smaller than the bandwidth!  This implies that the post-quench time evolution is sensitively dependent on even very weak disorder.  Some examples of this will be shown in Sec.~\ref{sec:time}.

\subsection{The continuum approximation to the post-quench occupation function}

In order to tell which features of the post-quench occupation function are due to the structure of the underlying lattice
and which, by contrast, are present also in a continuum treatment, 
we calculate the overlap $O_{qk}(\Delta j)$ for harmonic oscillator wavefunctions in the continuum; i.e.\ we compute the overlap of two harmonic oscillator eigenfunctions corresponding to the trap potential, one of which is displaced with respect to the other by $\Delta j$.
Some results on this continuum limit have already been obtained in \cite{Pezze2009}.

For convenience, we center the two eigenfunctions respectively at $x = \pm x_0$.  The overlap is given by:
\begin{equation}
\label{equ:overlapanalytic}
O_{qk}^{\text{cont.}}(x_{0})\equiv \int\limits_{-\infty}^{\infty}\psi^{*}_{q}(x +x_{0})\,\psi_{k}(x - x_{0})\,dx,
\end{equation}
 where the normalized harmonic oscillator eigenfunction is given by:
\be
\psi_{k}(x) = \frac{1}{\sqrt{2^{k} k!}}\pi^{-1/4}e^{-x^{2}/2}H_{k}(x).
\ee
Here $H_{k}(x)$ denotes the $k$th (physicists') Hermite polynomial, and we have chosen units in which $\hbar = m = \omega = 1$.
Using Eq.~(7.377) in \cite{Gradshteyn}, one finds:
\begin{equation}
\label{equ:overlapanalyticsol}
O_{qk}^{\text{cont.}} (x_0) =
\sqrt{\frac{2^\alpha \beta!}{2^\beta \alpha!}}\,(-1)^{{\rm
    max}(k-q,0)}\,e^{-x_0^2}\,x_0^{\alpha-\beta}\,L^{\alpha-\beta}_\beta(2x_0^2),
\end{equation}
where $\alpha \equiv \textnormal{max}(k,q)$, $\beta \equiv \textnormal{min}(k,q)$, and $L_{n}^{k}(x)$ are the associated Laguerre polynomials.

To complete our derivation we must relate the continuum shift of the eigenfunctions $x_0$ to the displacement of the harmonic trap in lattice units $\Delta j$.  The natural length scale of the continuum quantum harmonic oscillator is $\zeta = 1 / \sqrt{m\omega}$.
For the lattice problem, we may obtain expressions for $m$ and $\omega$ by Taylor-expanding the lattice kinetic energy $-2J \cos k$ around $k=0$.
This gives for the effective mass
\begin{equation}
m^{*}=\frac{1}{2J},
\end{equation}
while the effective frequency is given by
\begin{equation}
\label{equ:effomega}
\omega^{*} = \sqrt{\frac{\kappa}{m^{*}}} = \sqrt{2\kappa J}.
\end{equation}
The result is $\zeta = \left(2t/\kappa\right)^{1/4}$.
We thus find that $2 x_{0} = \Delta j/\zeta$.

\section{Time-evolution of experimental observables}
\label{sec:time}
\noindent
The occupation function analyzed in the previous section is the initial condition for the post-quench time-evolution of the atom cloud.
We now turn to the question of how this initial condition translates into the time-evolution of the cloud's spatial density profile.

The density of atoms at lattice site $j$ is given by the diagonal elements of the following equal-time Green's function:
\begin{equation}
\label{equ:correldef}
C_{ij}(t) \equiv \langle c^{\dagger}_{i}(t) c_{j}(t)\rangle.
\end{equation}
With a little algebra (see Appendix \ref{app:dens_calc}), we may write this in terms of the single-particle post-quench eigenfunctions and their eigenenergies.  This allows us to obtain the density profile at any time $t>0$:
\begin{equation}
\label{equ:density}
\rho_{j}(t) = \sum\limits_{l=1}^{N}\left\vert \sum\limits_{a=1}^{L} O_{al}e^{-iE_{a}t}\psi_{aj}\right\vert^2.
\end{equation}
Here $E_a$ is the eigenenergy of post-quench eigenstate $\beta_a$, and $\psi_{aj}$ is its (lattice) wave function.

The contributions of single-particle eigenstates $\beta_a$ and $\beta_b$ to post-quench observables dephase on a timescale $\tau_{ab} \sim 1/(E_a-E_b)$.  
This is largest for neighboring energy levels, $E_a$ and $E_{a+1}$.  This dephasing does not, of course, imply that the observables actually become time-independent, even at long times.  
However, if we examine an observable --- such as the density profile --- averaged over a time interval $\tau_{\rm av}$,
\begin{equation}
\label{equ:averagedensities}
\bar{\rho}_j (\tau_{\rm av}) \equiv \frac{1}{\tau_{\rm av}} \int\limits_0^{\tau_{\rm av}} \rho_j(t) dt,
\end{equation}
we find that this tends to a limiting form as $\tau_{\rm av} \to \infty$:
\begin{equation}
\label{equ:equilibriumdensities}
\bar{\rho}_j \equiv \lim_{\tau_{\rm av} \to \infty} \left( \bar{\rho}_j(\tau_{\rm av}) \right) =
\sum_{a=1}^L n_a \left\vert \psi_{aj} \right\vert^2.
\end{equation}
Following Deutsch \cite{Deutsch1991}, we call ${\bar \rho_j}$ the time-averaged density.

In the clean system (see Fig.~\ref{fig:ED_eigenstates}, upper panel), the density profile of every post-quench single-particle eigenstate is symmetric about the post-quench trap center $j_1$.
Hence the time-averaged density \eqref{equ:equilibriumdensities} will be centered at $j_1$ as well.
However, because the eigenenergies of the bonding and anti-bonding Bragg-localized states are very nearly degenerate, the restoration of this symmetry about $j=j_1$ occurs very slowly.  This is demonstrated in Fig.~\ref{fig:CoM_Skew}, where the cloud's center of mass seemingly equilibrates at a position between the original trap center $j_0$ and the new trap center $j_1$.  In reality, though, a very slow drift --- not visible on experimental timescales --- will eventually restore the center of mass to $j=j_1$ (see Fig.~\ref{fig:dens_distr_thermal}).

However, this symmetry-restoring drift ceases to occur as soon as the disorder is able to disrupt the hybridization between the left- and right-localized states.  As discussed above, this occurs for any $W \gtrsim T$, where $T$ --- given in (\ref{equ:matelT}) --- is exponentially small in $2j_b(E)$, the width of the Bragg-forbidden region.
Therefore, even for such weak disorder, the parity-breaking imposed by the initial conditions remains visible in the infinite-time density profile (see Fig.~\ref{fig:dens_inf_dis}).  This is a localization mechanism for the atom cloud which is conceptually quite distinct from Anderson localization.

As a diagnostic for this we define the jump efficiency, which expresses the post-quench displacement of the center of mass as a fraction of the jump size $\Delta j$. 
Fig.~\ref{fig:jump_eff} shows a plot of the jump efficiency as a function of jump size for various disorder strengths.  This clearly demonstrates the distinction between Bragg and Anderson localization.

In the remainder of this section we will study two facets of the post-quench density profile --- its early-time behavior and its time-averaged value --- in more detail.
The early-time behavior, analyzed in Sec.~\ref{subsec:earlytime}, is similar for the clean and weakly disordered cases.  The time-averaged state, however, is not; therefore, we analyze the clean case in Sec.~\ref{subsec:longtimeclean}, and then the disordered cases in Sec.~\ref{subsec:longtimedirty}.

\subsection{Early-time behavior}
\label{subsec:earlytime}
\begin{figure}
\centerline{\includegraphics[width=\columnwidth]{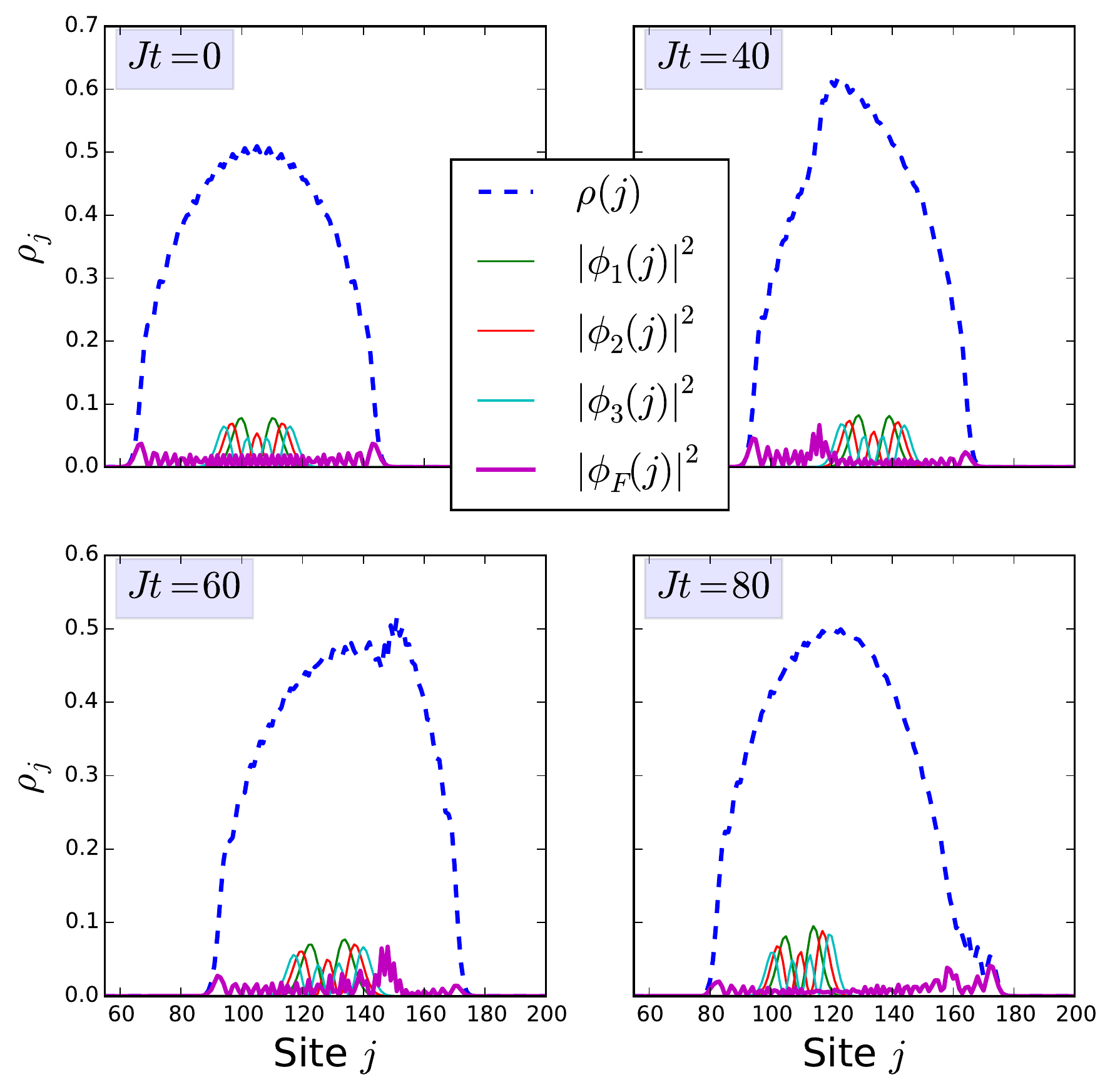}}
\caption{The density profile of the atom cloud at various times after the quench (blue dashed curve), showing the contributions of selected individual pre-quench single-particle eigenstates (solid curves, various colors).  The skewness oscillations are caused by the mobile `bump' in the profile, which lags behind the center-of-mass oscillations, and which appears to be due principally to the highest-lying occupied eigenfunction $\phi_F(j)$.
Parameters:\ number of lattice sites $L = 241$; trap spring constant $\kappa = 0.0025$; hopping integral $J = 1$; chemical potential $\mu =0$; pre-quench trap center $j_{0} = 106$; post-quench trap center $j_{1} = 121$; disorder strength $W = 0$.}
\label{fig:dens_time}
\end{figure}

To characterize the time-evolution of the density \eqref{equ:density} shortly after the quench, we consider in particular two of its moments:\ the first moment, $x_1$, which corresponds to the atom cloud's center of mass; and the third (standardized) moment, $x_3$, which corresponds to its skewness.
These are defined respectively as
\begin{equation}
\label{equ:COM}
x_{1}(t) \equiv \frac{1}{N}\sum\limits_{j=1}^{L} j \,\rho_{j}(t)
\end{equation}
and
\begin{equation}
\label{equ:Skewness}
x_{3}(t) \equiv  \frac{\frac{1}{N}\sum_{j}\left(j-x_{1}(t)\right)^{3}\rho_{j}(t)}{\left(\frac{1}{N}\sum_{j}\left(j-x_{1}(t)\right)^{2}\rho_{j}(t)\right)^{3/2}}.
\end{equation}
We plot them as functions of time in Fig.~\ref{fig:CoM_Skew}, for two different jump sizes.

The dominant effect is clearly the oscillation of the center of mass, the frequency of which may be accurately predicted by a classical oscillator calculation using the band mass as the mass of the particle --- see \eqref{equ:effomega}.  In this case, a spring constant of $\kappa = 0.0025$ and a hopping integral of $J = 1$ yield a frequency of $f/J = 0.01125$, which matches the oscillation frequency of $x_1$ in Fig.~\ref{fig:CoM_Skew}.

For the smaller jump size, this oscillation occurs about the post-quench trap center, $j_1$, which is shown by the solid horizontal (black) line.  For the larger jump size, however, it appears to occur around a different point, somewhere between $j_0$ and $j_1$.  As discussed above, this is because the quench with the larger jump size populates some of the left Bragg-localized states, which on the timescales shown have not yet tunneled across to their partners on the right.

The oscillations in the skewness are much smaller-scale than those of the center of mass.
In Fig.~\ref{fig:dens_time}, we elucidate their origin by plotting the contributions of selected individual single-particle eigenfunctions to the overall density profile.
This decomposition of the density strongly suggests that the skewness oscillation is a finite-size effect.
This is supported by exact diagonalization for larger values of the chemical potential, which suggests that the skewness oscillations are suppressed as $N$ increases, and also by the solution of the fermionic Gross-Pitaevskii equation \cite{Kim2004}, which suggests that they are absent in the continuum.
Nonetheless, for typical experimental set-ups, in which one may have $N \sim 100$ atoms per quasi-one-dimensional tube, they may well be observable.

\subsection{Time-averaged state (clean case)}
\label{subsec:longtimeclean}
\begin{figure}
\centerline{\includegraphics[width=0.95\columnwidth]{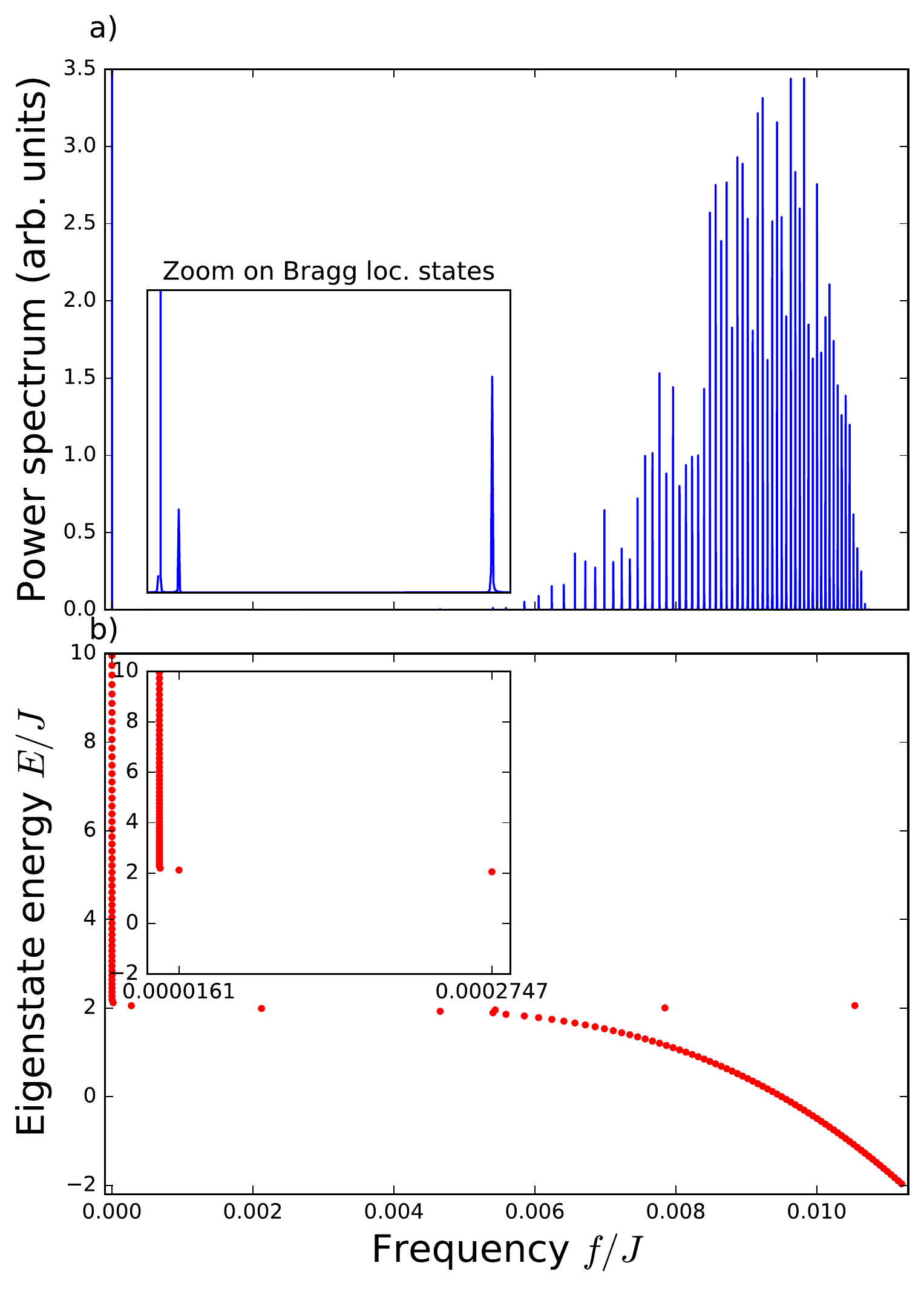}}
\caption{The correspondence between the long-time part of the center-of-mass oscillations and the dephasing of nearly degenerate Bragg-localized states.
(a) The power spectrum of the center-of-mass oscillations of the atom cloud.
(b) Energy differences of neighboring single-particle energy levels converted to frequencies (horizontal axis) for a pair of single-particle states near energy $E$ (vertical axis).
Insets:\ a zoomed-in version of the same, showing the first two Bragg-localized states.
Note the excellent quantitative match between the frequency content of the upper and lower panels.
Parameters:\ number of lattice sites $L = 241$; trap spring constant $\kappa = 0.0025$; hopping integral $J = 1$; chemical potential $\mu =0$; pre-quench trap center $j_{0} = 105$; post-quench trap center $j_{1} = 121$; disorder strength $W = 0$.}
\label{fig:power_spectrum_ana}
\end{figure}

\begin{figure}
\centerline{\includegraphics[width=\columnwidth]{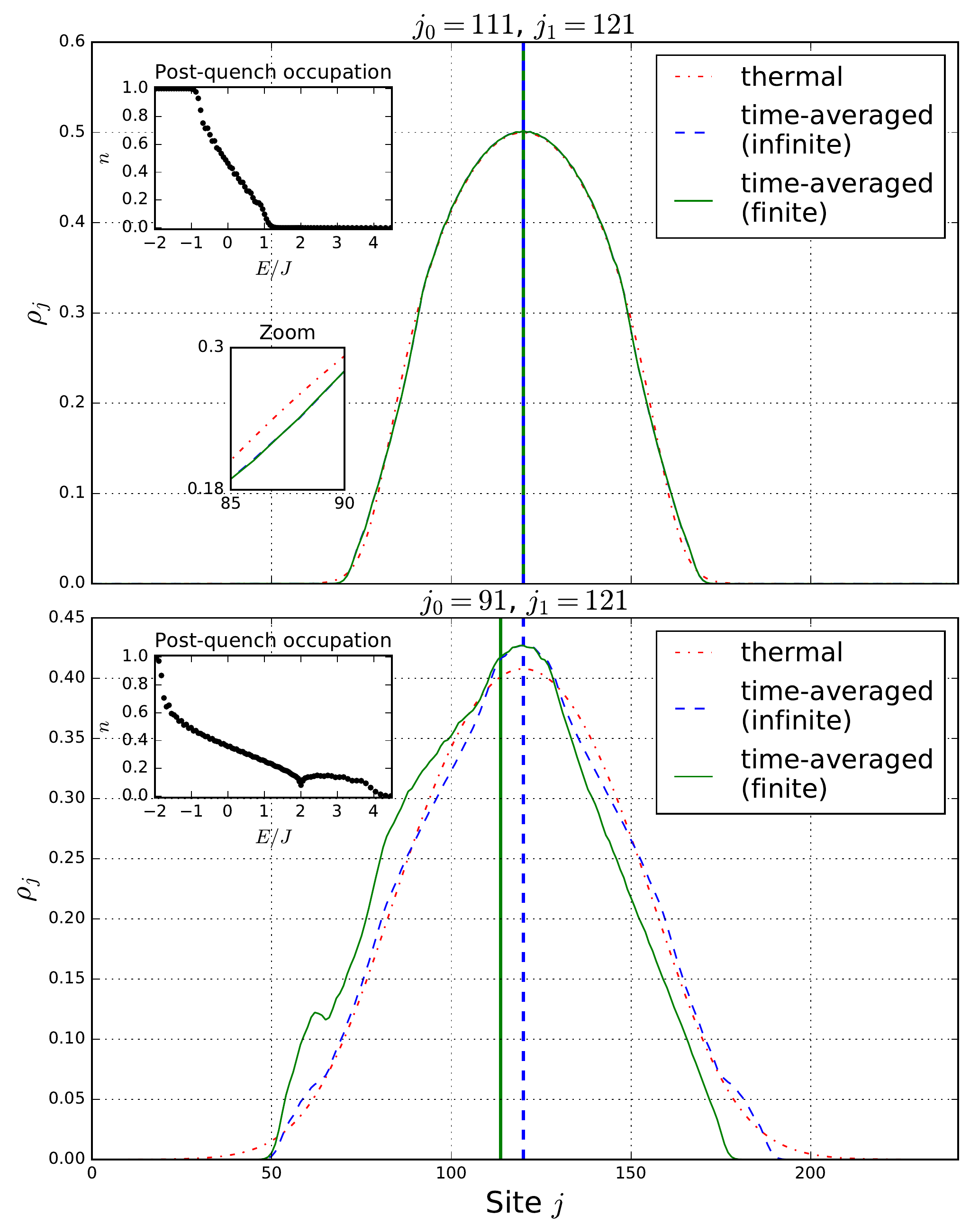}}
\caption{A comparison between three densities:\ the time-averaged density calculated from expression (\ref{equ:equilibriumdensities}) (`time-averaged (infinite)'); the time-average of consecutive densities at large but finite times (`time-averaged (finite)') (similar to \eqref{equ:averagedensities}); and the density of a thermal equilibrium state with the same total energy and number of atoms (`thermal').
The finite average was taken at $Jt = 50000$ for $1000$ consecutive time steps separated by $\Delta t = 1/J$.
The vertical lines denote the position of the center of mass for the corresponding density.
The time-averaged density does not perfectly match the thermal prediction for any non-zero jump size.
For larger jump sizes (bottom), as the Bragg-localized states are populated, the approach to the time-averaged state becomes very slow. 
This happens because Bragg localization generates a very long time-scale, Eq.~\ref{equ:longtimescale}, below which a time-average deviates strongly from the infinite-time result.
This is shown by the disparity between the `time-averaged (finite)' and `time-averaged (infinite)' curves in the lower panel.
Parameters:\ number of lattice sites $L = 241$; trap spring constant $\kappa = 0.0025$; hopping integral $J = 1$; chemical potential $\mu =0$; disorder strength $W = 0$.}
\label{fig:dens_distr_thermal}
\end{figure}

As emphasized above, in the clean case all single-particle eigenstates have densities symmetric about the post-quench trap center, which means that the time-averaged density profile will have this symmetry too.  Therefore, we should be able to see in the power spectrum of the center-of-mass oscillations the slow modes that restore this symmetry at long times.
As shown in Fig.~\ref{fig:power_spectrum_ana}, indeed we can.  Panel (a) shows the frequencies present in the power spectrum, with an inset concentrating on the low-frequency spectrum.  Panel (b) is a histogram of the frequencies obtained from the gaps between neighboring post-quench single-particle energy levels.  The quantitative match between these graphs is striking.
Furthermore, the oscillation frequency calculated above ($f/J = 0.01125$) provides an upper bound to the frequency spectrum.

This analysis demonstrates how long a time scale one would need to go to to see the atomic cloud oscillating about the new trap center.
This time may be estimated as the dephasing time of the highest occupied Bragg-localized state, i.e.
\be
\label{equ:longtimescale}
\tau_{\textnormal{long}}\approx 1/T_{F} = e^{j_{b}(E_F)/\xi(E_F)},
\ee
where $E_F$ denotes the eigenenergy of that state, and the functions $j_b(E)$ and $\xi(E)$ are defined in (\ref{jbe}) and (\ref{xie}) respectively.
This should be compared with the time scale associated with the center-of-mass oscillations immediately after the quench, which is given by
\be
\tau_{\text{short}}\approx 1/f^* = \frac{2 \pi}{\sqrt{2\kappa J}}.
\ee
Only for times $\tau_{\rm av} \gg \tau_{\rm long}$ will a time-averaged density profile match the symmetric prediction of \eqref{equ:equilibriumdensities}.

While this time-averaged density has a center of mass which matches the thermal equilibrium prediction, other moments of the time-averaged and thermal profiles do not agree, as shown in Fig.~\ref{fig:dens_distr_thermal}.
Due to the reflection symmetry about $j=j_1$, the center of mass, the skewness, and in fact all odd moments of the density do `thermalize'.  However, the same is not true for the even moments:\ even for small trap jumps, the two densities are different.
Related questions have also been discussed for hard-core bosons~\cite{Rigol2006}.

In addition to the two densities obtained from the occupation functions, we have plotted an average density over many consecutive time steps at very large times.
This underlines that (a) there is a long period of time over which the density reaches a `finite-time-averaged' state, where the in-band single-particle states have dephased but the weakly-hybridized 
pairs of Bragg-localized states have not, and (b) the true time-averaged density emerges only at significantly longer times than used in this example.

\subsection{Time-averaged state (disordered case)}
\label{subsec:longtimedirty}
\begin{figure}
\centerline{\includegraphics[width=\columnwidth]{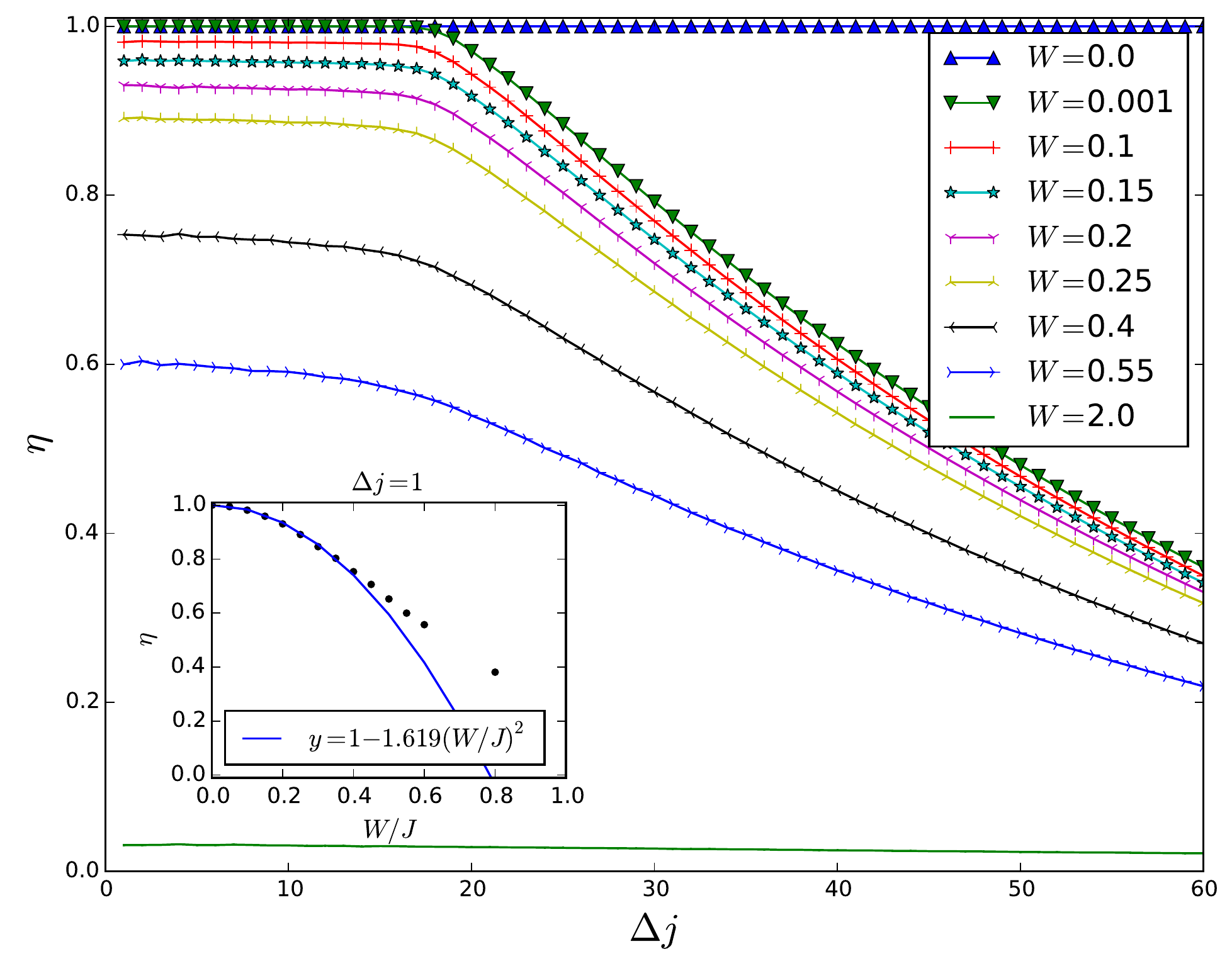}}
\caption{The jump efficiency $\eta$ as a function of the jump size $\Delta j$, for various disorder strengths.  This graph illustrates the qualitative distinction between Anderson and Bragg localization.
For $W = 0$ the jump efficiency $\eta$ is always unity.
For the disordered cases Bragg localization appears in the form of a $\Delta j$-dependent decrease in $\eta$.
Inset: The jump efficiency $\eta$ as a function of disorder strength $W/J$ for a fixed jump size $\Delta j = 1$.
Parameters:\ number of lattice sites $L = 241$; trap spring constant $\kappa = 0.0025$; hopping integral $J = 1$; chemical potential $\mu = 0$; post-quench trap center $j_{1} = 121$.
Each disorder-average is performed over 10000 disorder realizations.}
\label{fig:jump_eff}
\end{figure}

As we have already emphasized, we find that even weak disorder, provided that it is large compared to the splitting between symmetric and anti-symmetric Bragg-localized states, causes the time-averaged density to be significantly asymmetric about the new trap center.  
This asymmetry, which would be impossible in thermal equilibrium, can be seen in Fig.~\ref{fig:dens_inf_dis}.

The reason for the asymmetry is twofold.  First, an arbitrarily weak disorder potential breaks the parity symmetry of the clean Hamiltonian.  
This has the consequence, for $W \gtrsim T$, that the eigenstates become localized on the left or the right of the trap.
Second, as disorder is made stronger, this effect extends to the delocalized states in the center of the trap.

To quantify the influence of disorder, we define the `jump efficiency' $\eta$ as follows:
\begin{equation}
\label{equ:jumpefficiency}
\eta \equiv \frac{x_{1}^{t \rightarrow \infty}-x_{1}^{t =0}}{\Delta j}.
\end{equation}
Here the pre-quench center-of-mass position of the cloud, $x_1^{t=0}$, is calculated from the pre-quench distribution; the time-averaged post-quench center-of-mass position, $x_{1}^{t \rightarrow \infty}$, is calculated from \eqref{equ:equilibriumdensities}.  
Put simply, this jump efficiency describes (as a number between 0 and 1) how much of the way from the pre-quench trap center to the post-quench trap center the atom-cloud moves.

Fig.~\ref{fig:jump_eff} shows the jump efficiency as a function of jump size for different disorder strengths.  
The most striking feature is that, even in the limit where the jump size $\Delta j \to 0$, the jump efficiency does not remain unity; rather, it has the form
\begin{equation}
\eta_{\rm p} \equiv \lim_{\Delta j \to 0} \left[ \eta(\Delta j) \right] \approx 1 - \alpha W^2. \label{etaplat}
\end{equation}
This may be understood as the development of a correlation between (a) whether the disorder potential shifts the center-of-mass of a particular post-quench eigenfunction to the left or to the right, and (b) whether the post-quench occupation of that eigenfunction goes up or down. 
Each of these effects is first-order in the disorder potential $V(x)$, but each by itself would average to zero. 
However, the development of a correlation between them gives an effect of order $W^2$ that survives the disorder average. 
We present in appendix \ref{app:jumpeff} a toy calculation that displays this physics.

As the jump size is increased, the plateau in $\eta(\Delta j)$ at some point gives way to a decrease in the jump efficiency. 
This is because the jump size is now large enough to populate some of the Bragg-localized states of the post-quench trapping potential. 
As discussed above, these are sensitive to even weak disorder, and once localized they effectively contribute almost nothing to the jump efficiency. 
Indeed, the shape of the curves in Fig.~\ref{fig:jump_eff} may be well approximated by the equation
\begin{equation}
\eta = f_{\rm d} \eta_{\rm p},
\end{equation}
where $f_{\rm d}$ is the fraction of the pre-quench atoms that are projected into non-Bragg-localized states, and $\eta_{\rm p}$ is the `plateau value' of the jump efficiency defined in (\ref{etaplat}).
As previously stated, to obtain the Bragg-localization effects visible in Fig.~\ref{fig:jump_eff} we require only a parity-breaking potential while features associated with Anderson localization require randomness.

\section{Conclusion}
\label{sec:conclusion}
\noindent
We have studied a particular type of relatively simple quantum quench:\ a sudden trap displacement applied to a one-dimensional system of non-interacting lattice fermions with disorder. 
The central theme of this work is to provide an understanding of how confinement, lattice structure and disorder conspire to provide
various dynamical regimes to the coherent post-quench time evolution. 
We discuss these questions using a number of relatively straightforward real-space observables. 

Our main observation is that the disorder in this system has two distinct localizing effects:\ Anderson localization, which occurs via the same mechanism as in the untrapped system, and Bragg localization, which arises from the presence in the single-particle spectrum of the post-quench Hamiltonian of nearly degenerate bonding and antibonding states that are spatially localized near the edges of the trap.

As a result of Bragg localization, the time-evolution of the density profile of the clean system after a quench shows two regimes. 
In the short-time regime, the dynamics are driven by the dephasing of the `in-band' states (those with energies $\vert E \vert < 2J$), and look like collective oscillations about a position which may not match that of the actual post-quench trap center. 
In the long-time regime, the dephasing of the Bragg-localized states causes a slow drift of the center of mass from this position to the center of the trap.

The role of disorder in the long-time evolution is very pronounced. 
Since the splitting between  the symmetric and antisymmetric combinations
of the Bragg-localized states is exponentially suppressed in their separation, 
extremely weak disorder can dominate over this splitting, resulting in 
a time-averaged state which magnifies the weak parity-breaking of the disorder potential into a macroscopic effect.  
Indeed, for the system parameters we have studied, as seen in Fig.~\ref{fig:jump_eff}, a disorder strength of 
less than a thousandth of the bandwidth of the single-particle hopping band can reduce the jump efficiency by a factor of more than two!

We have assumed throughout that the Bragg-localized level pairs form a discrete spectrum, and in that sense all of this analysis is for a finite-size system. This is the case for which  experiments are perhaps most likely to be realised initially.  
However, it is interesting to ask what would happen if we took the thermodynamic limit.  
Then the disorder potential would make one left-Bragg-localized state resonant with a different right-Bragg-localized state.  
Would this still suppress the jump efficiency from unity?  If so, by what fraction?

It would also, of course, be interesting to consider the introduction of interactions between the fermions.  
This would allow us to investigate, for example, whether the logarithmic growth of entanglement entropy seen in many-body-localized systems also occurs when interactions are added to the Bragg-localized case. 
Another related question is whether interactions naturally counteract Bragg localisation. Such an analysis, carried out by time-evolving block decimation (TEBD), is underway \footnote{M. Schulz, F. Pollmann, C.A. Hooley, and R. Moessner, in preparation}. 

Overall, we believe that this kind of  quantum quench  provides an ideal platform for studying the interplay of spatial inhomogeneity, disorder and interactions 
for the dynamics in a quantum coherent setting.

\section{Acknowledgments}
\label{sec:ack}
\noindent
MS acknowledges support from EPSRC (UK) via the CM-CDT program, grant number EP/L015110/1.
CAH's work on this paper was performed in part at the Aspen Center for Physics, which is supported by National Science Foundation grant PHY-1066293.  
He is grateful to them for their hospitality.  
He is also thankful for ongoing support from the EPSRC (UK) via the TOPNES program, grant number EP/I031014/1.

\appendix
\section{The zero-hopping case}
\label{app:occ_hop_dis}

\begin{figure}
\centerline{\includegraphics[width=1.0\columnwidth]{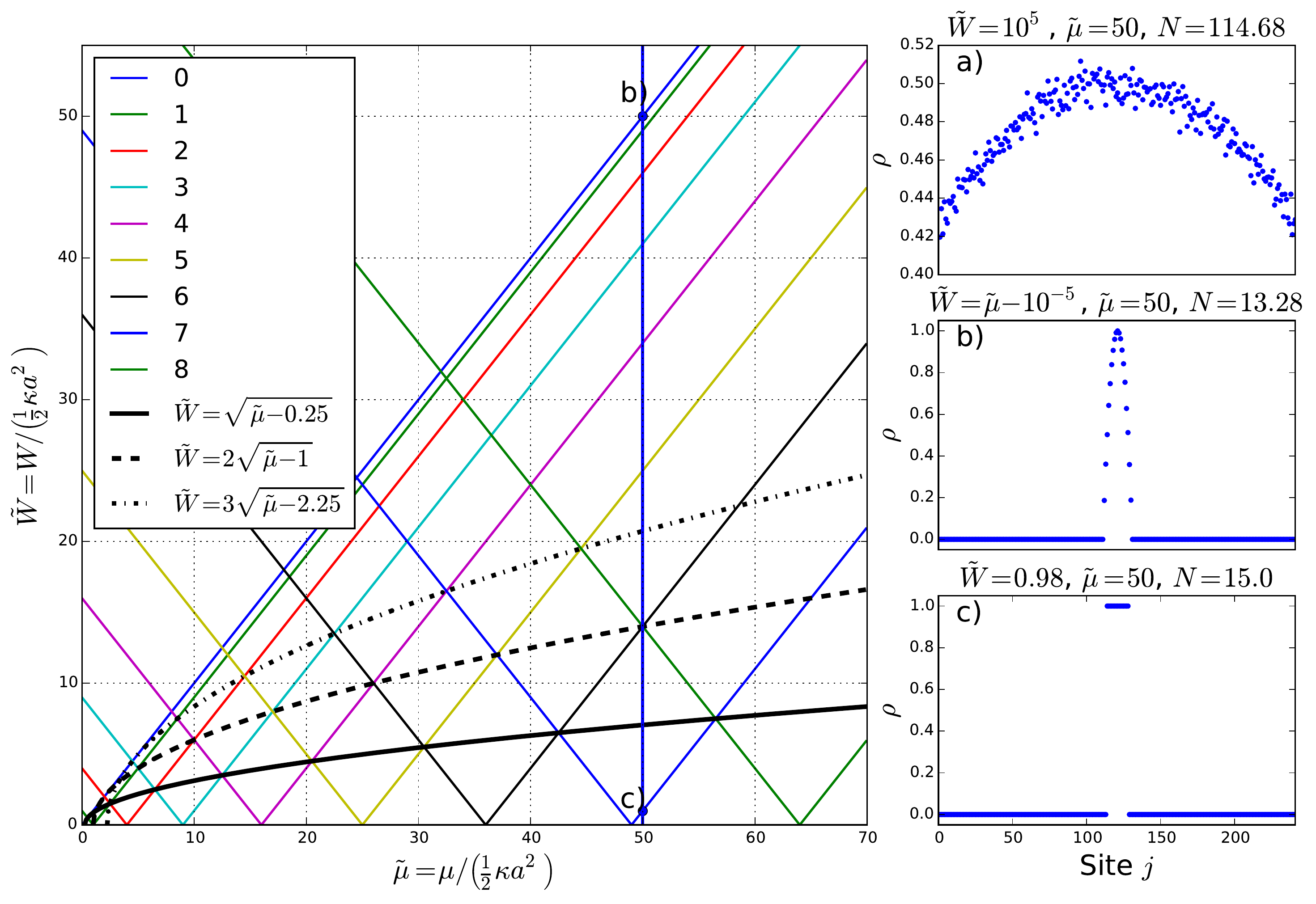}}
\caption{Left panel: An illustration of the disorder-broadening of the single-particle energy levels of a finite-size system. Possible level crossings as a function of $\tilde{W}$ and $\tilde{\mu}$ lie on parabolic curves showing the critical disorder strength at which neighboring (black solid line), second-neighbor (black dashed line), third-neighbor (black dotted line), etc.\ levels can cross.
Right panels: The disorder-averaged density profile for three different disorder strengths.
Parameters:\ number of lattice sites $L = 241$; trap spring constant $\kappa = 0.0025$; hopping integral $J = 0$; trap center $j_{0} = 121$.}
\label{fig:density}
\end{figure}

\begin{figure}
\centerline{\includegraphics[width=\columnwidth]{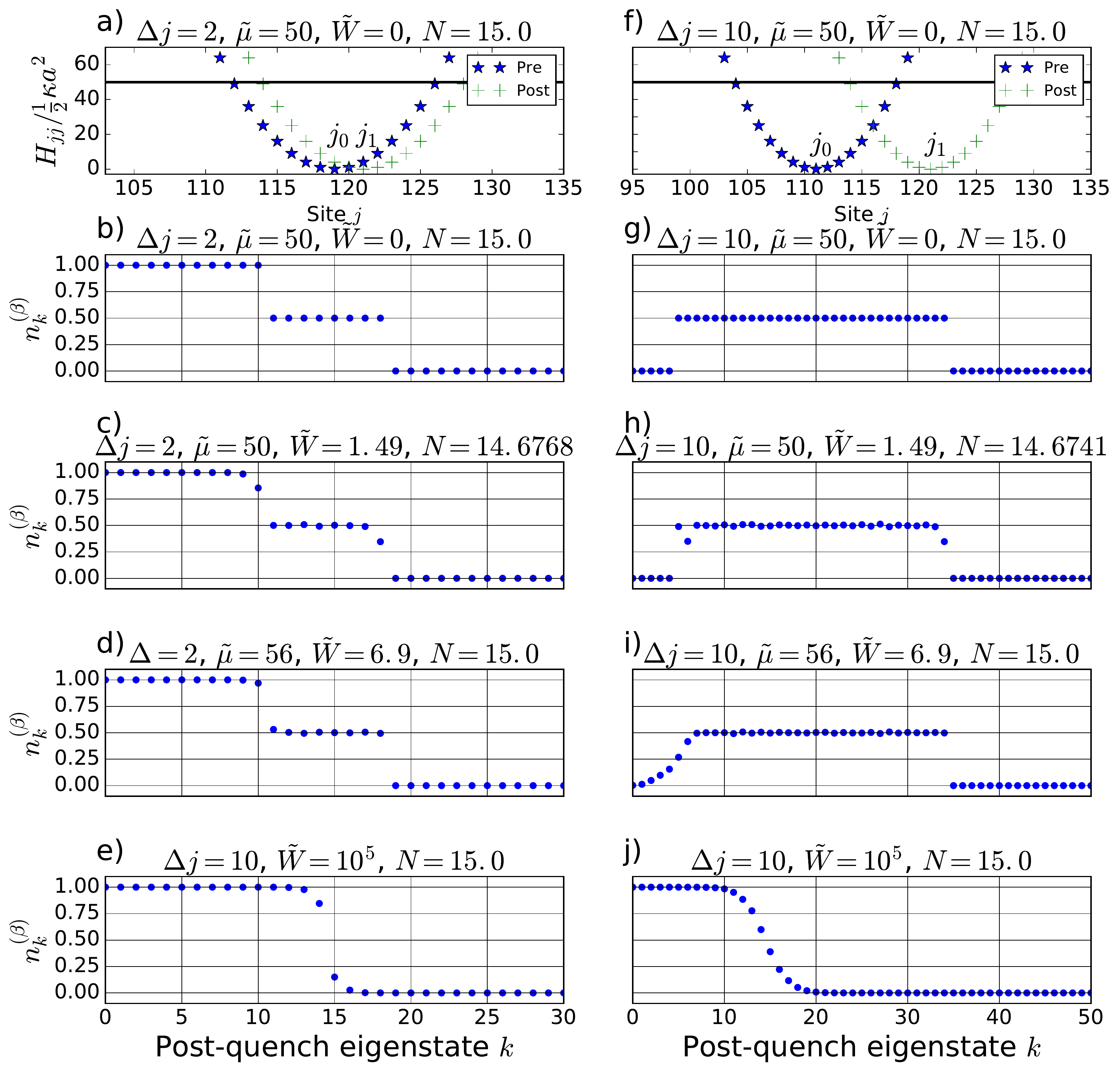}}
\caption{An illustration of the different qualitative forms of the post-quench occupation function in the absence of hopping ($J=0$), for different jump sizes $\Delta j$, scaled chemical potentials $\tilde{\mu} \equiv 2\mu/\kappa a^2$, and scaled disorder strengths, $\tilde{W} = 2W/\kappa a^2$.
(a) The pre- and post-quench trapping potentials.  In the absence of hopping and with no disorder, this is also a graph of the eigenvalues of the pre- and post-quench Hamiltonians.  The black line denotes the chemical potential $\tilde{\mu}$.
(b) The post-quench occupation function after a small trap-jump, as a function of eigenstate quantum number $k$.
(c) The same as (b), but with moderate disorder.  The deviation from half-unit values is mainly because changing the disorder at fixed chemical potential changes the total particle number.
(d) The same as (c), but with a choice of $\tilde{W}$ and $\tilde{\mu}$ that restores the original particle number $N$.  Note that there are still residual deviations from half-unit values.
(e) The occupation function in the case of very large $\tilde{W}$.
(f)-(j) As panels (a)-(e), but for a larger jump size, $\Delta j =10$.
Parameters:\ number of lattice sites $L = 241$; trap spring constant $\kappa = 0.0025$; hopping integral $J = 0$; pre-quench trap center $j_{0} = 121$.
Each disorder-average is performed over 10000 disorder realizations.}
\label{fig:distr_nohop}
\end{figure}
\noindent
As an aid to understanding the disorder-dependence of the pre-quench density $\rho(t=0)$ and the post-quench occupation function $n^{(\beta)}_{k}$, we discuss in this appendix the form they take in the zero-hopping ($J=0$) case.  Without any hopping, the pre- and post-quench eigenfunctions can be chosen to be eigenfunctions of position.  However, in order to connect smoothly to the $J \ne 0$ case, we instead take the excited states to be bonding and antibonding superpositions of the pair of parity-related degenerate position eigenstates.

The left panel of Fig.~\ref{fig:density} shows the disorder-broadening of the discrete energy levels of the harmonic trap.
It permits us to determine possible level crossings and level occupations for given values of the chemical potential $\mu$ and the disorder strength $W$.
The parabolic curves show the disorder strength at which neighboring (next-neighbor, etc.)\ levels first  cross.
However, the only line of relevance in the continuum case is in fact the line $W=\mu$, which denotes the broadening of (what was in the clean case) the single-particle ground state.
This divides the $(W,\mu)$ parameter space into two distinct regions.

The right panels illustrate the qualitative difference in the form of the ground-state density between these two regions.  Panel (a) shows a case where $W \gg \mu$.  Here all lattice sites are occupied with roughly equal probabilities, though the breaking of particle-hole symmetry due to the trapping potential is still visible.  Panel (c) shows a contrasting case where $W \ll \mu$.  Here the spatial density profile has a `top hat' form.  Panel (b) shows the density at the point $W = \mu$:\ here the average occupation of (what was in the clean case) the lowest energy site is just about to deviate from unity.

A careful analysis of Fig.~\ref{fig:density} is necessary to understand the occupation function obtained in the $J = 0$ quench problem, some examples of which are shown in Fig.~\ref{fig:distr_nohop}.
Due to the lack of hopping, in a single disorder realization the density can only take the values 0 or 1.  The same is true of the occupation function --- except in the clean case, where our choice of bonding and antibonding forms of the eigenstates allows also a value of 1/2.  The disorder-averaging, of course, permits other values to emerge as weighted averages of these.

In panels (a) and (f) we visualize the quench protocol by showing the diagonal matrix elements (which for zero hopping are also the eigenvalues) of $\hat{H}_{i}$ and $\hat{H}$.
The translation of the trap explains the shape of the disorder-free occupation function for the different jump sizes in panels (b) and (g).  These are in the case $W \ll \mu$, so the real-space pre-quench density is of top-hat form, i.e.\ just one continuous block of occupied sites.  Where the two sites corresponding to a degenerate pair of post-quench eigenstates both exist within that block, those states get occupation 1; where only one of the sites overlaps with the original density profile, they get occupation 1/2; and where both sites lie outside the block, they get occupation 0.

Adding disorder to the system allows the levels to cross, and also leads, for a fixed chemical potential, to a change of the total particle number.
This means that disorder distorts the clean post-quench distribution function in two qualitatively different ways.  These are shown separately in panels (c), (d), (h), and (i).

Panels (c) and (h) show the occupation function when $\mu$ and $W$ are chosen so as not to mix any neighboring levels (i.e.\ below the thick black parabola in Fig.~\ref{fig:density}).  However, different disorder realizations may still push the highest occupied level through the chemical potential, resulting in an average total particle number that is non-integer.

In panels (d) and (i) the disorder is strengthened, but the chemical potential is also raised.  This results in the opposite situation:\ now the disorder cannot empty a previously occupied state, but there is on the other hand a strong possibility of the lower levels' being permuted.
Since the energy-level permutation is more likely at lower energies, the departure from the clean behavior is asymmetric, unlike in panels (c) and (h).

Lastly, we have included the case of very strong disorder, for comparison with Fig.~\ref{fig:distr_hop_multdis}.

\section{The post-quench occupation function with moderate disorder}
\label{app:hopping-and-disorder}

\begin{figure}
\centerline{\includegraphics[width=\columnwidth]{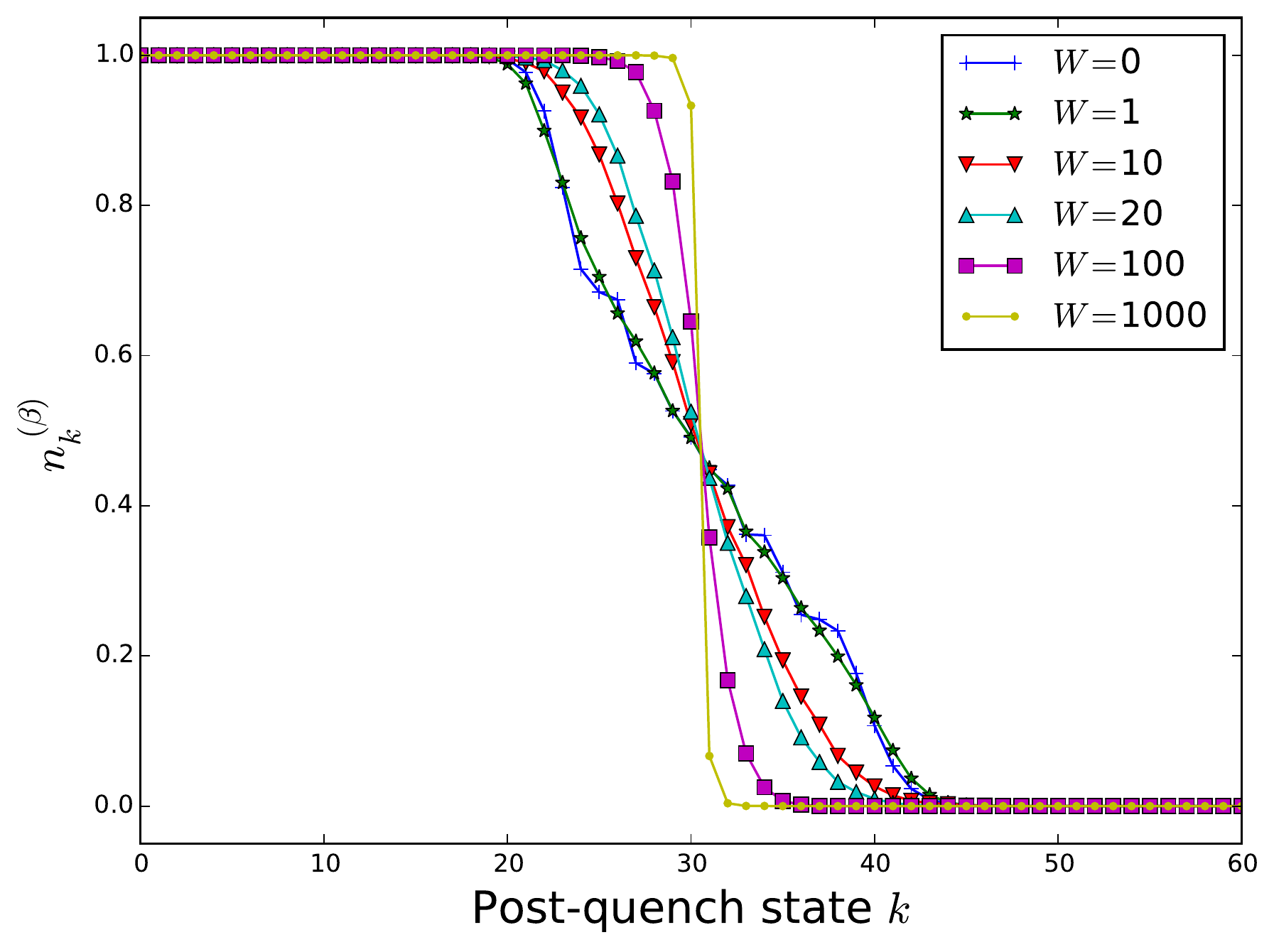}}
\caption{The post-quench occupation function $n_k^{(\beta)}$ in the presence of hopping $J$ and moderate disorder $W$.
For convenience we have picked a constant particle number $N=31$.
Parameters:\ number of lattice sites $L = 241$; trap spring constant $\kappa = 0.0025$; hopping integral $J = 1$; pre-quench trap center $j_0=115$, post-quench trap center $j_{1} = 121$.
Each disorder-average is performed over 10000 disorder realizations.}
\label{fig:distr_hop_multdis}
\end{figure}

\noindent
In addition to the zero-hopping case, we have analyzed what happens to the occupation function when we introduce moderate disorder into the $J \ne 0$ system.  (Here `moderate' means a disorder strength high enough to do more than just lift the degeneracy between neighboring Bragg-localized states.)
The results of this analysis are shown in Fig.~\ref{fig:distr_hop_multdis}.
In order to have comparable results upon disorder averaging we have chosen a fixed particle number $N$ rather than a fixed chemical potential $\mu$.
We have plotted the occupation function not as a function of energy, but rather eigenstate quantum number $k$ ordered by energy.

The results show that as the disorder becomes stronger, the trap jump has an increasingly minor effect upon the post-quench occupation function.
This is as expected, since the disorder profile, unlike the trapping potential, is not displaced at the moment of the quench.
Furthermore we see that in the case of very strong disorder the occupation functions for the $J = 0$ model (see Fig.~\ref{fig:distr_nohop}) and that for the $J \neq 0$ model become qualitatively similar. 
This is again as it should be, since the nearest-neighbor coherence $\left\vert\left\langle J c^{\dagger}_{i}c_{i+1}\right\rangle \right\vert \approx \frac{J^2}{W}$ for large disorder.

\section{Hybridization between left- and right-Bragg-localized states}
\label{app:MatrixT}
\noindent
In this appendix we obtain an approximate form for the matrix element $T$ responsible for the hybridization of left- and right-Bragg-localized states.  The calculation is similar in structure to that of the hopping integral in a tight-binding model.

We first split the Hamiltonian into three parts:
\begin{equation}
H = H_{\text{kin}}+V_0 + V_1.
\end{equation}
Here $H_{\rm kin}$ is the lattice kinetic energy,
\be
H_{\rm kin} = - 2J \cos ( {\hat k} a ),
\ee
while the potential terms $V_0$ and $V_1$ are defined as follows:
\bea
V_0(x) & = & \frac{1}{2} \kappa x^2 \,\Theta(-x), \\
V_1(x) & = & \frac{1}{2} \kappa x^2 \,\Theta(x),
\eea
where $\Theta(x)$ is the step function.  $H_L \equiv H_{\rm kin} + V_0$ has only left-localized eigenstates, while $H_R \equiv H_{\rm kin} + V_1$ has only right-localized ones.  We may thus calculate the hopping integral from the left- to the right-localized states by introducing $V_1$ as a perturbation to $H_L$.

Following \citep{Hooley2004}, we use a WKB approximation for the left-localized eigenstate, i.e.\ an eigenstate of $H_L$ with eigenenergy $E$:
\be
\phi_L(x) \sim \exp\left( i \int\limits_{x_0}^x k(x') dx' \right),
\ee
where the wavenumber $k(x)$ is the solution to the equation
\be
-2J \cos (ka) + V_0(x) = E
\ee
and $x_0$ is an arbitrary reference point.  We see that, for $x>0$, $k$ is independent of $x$.  The calculation is not very sensitive to the structure of $\phi_L(x)$ for $x<0$, so we make the following rather crude approximation:
\begin{equation}
\phi_L(x) = \begin{cases}
\displaystyle   0 & x\leq -x_{c}; \\
& \\
\displaystyle    \frac{1}{\sqrt{x_{c}(E)-x_{b}(E)+\xi(E)}}~ & -x_{c}<x\leq -x_{b}; \\
& \\
\displaystyle    \frac{e^{-x/\xi(E)}}{\sqrt{x_{c}(E)-x_{b}(E)+\xi(E)}}~ & x > -x_{b}.
  \end{cases}
\end{equation}
Here $x_c(E)$ and $x_b(E)$ are respectively the classical and Bragg turning points of the semiclassical orbit,
\be
x_c(E) = \sqrt{\frac{2E+4J}{\kappa}}, \qquad
x_b(E) = \sqrt{\frac{2E-4J}{\kappa}},
\ee
and $\xi(E)$ is the decay length in the Bragg-forbidden region,
\be
\xi(E) = -\frac{a}{\text{ln}\left(\frac{E}{2J}-\sqrt{\left(\frac{E}{2J}\right)^2-1}\right)}.
\ee
Since the transformation $x \to -x$ transforms $H_L$ into $H_R$, it follows that the eigenstate of $H_R$ with energy $E$ is given by $\phi_R(x) = \phi_L(-x)$.

The hopping integral is
\be
T = \int\limits_{-\infty}^\infty \phi_L^{*}(x) \, V_1(x) \,\phi_R(x) \,dx.
\ee
This integral is dominated by the region in which $\phi_R(x)$ is constant; hence
\be
T \approx \frac{\kappa}{2} \int\limits_{x_b(E)}^{x_c(E)} \frac{x^2 \, e^{-x/\xi(E)}}{x_{c}(E)-x_{b}(E)+\xi(E)} dx.
\ee
For large energies we can approximate this integral as:
\begin{equation}
T(E) \approx \frac{\kappa}{2} \left[ x_b(E) \right]^2 \exp \left( - \frac{x_{b}(E)}{\xi(E)} \right),
\end{equation}
which is the form quoted in (\ref{equ:matelT}).

\section{The equal-time Green's function}
\label{app:dens_calc}
\noindent
In this appendix we present a general derivation of the form of the equal-time Green's function (\ref{equ:correldef}).  This is useful for our purposes because its diagonal form gives the density; but it may also be useful in future work for calculating such quantities as the entanglement entropy \cite{Peschel2012}.

Defining the following basis transformations:
\begin{eqnarray}
c_{j} & = & \phi^{*}_{kj}\alpha_{k}; \\
c_{j} & = & \psi^{*}_{kj}\beta_{k}; \\
\beta_{k} & = & O_{kq} \alpha_{q},
\label{equ:transformations}
\end{eqnarray}
we can expand \eqref{equ:correldef} to
\begin{equation}
C_{ij}(t) = \langle \psi_0^{(N)}\vert e^{iHt}\phi_{ki}\phi^{*}_{qj}\alpha_{k}^{\dagger}\alpha_{q}e^{-iHt}\vert \psi_0^{(N)}\rangle. \label{equ:twoptF}
\end{equation}
(Here and in the rest of this appendix we use the Einstein convention that repeated indices are summed over.)  From Blaizot's book \cite{Blaizot1986} (2.19) we get the following identity:
\begin{equation}
\label{equ:Blaizot}
e^{\frac{1}{2}\tilde{\nu}K\nu+\bar{l}\nu}\nu_{i}e^{-\frac{1}{2}\tilde{\nu}K\nu-\bar{l}\nu}=\sum\limits^{2L}_{j=1}\left[\left(e^{-\sigma K}\right)_{ij}\nu_{j}+l_{j}\sigma_{ji}\right],
\end{equation}
where $\tilde{\nu} = \nu^{\textnormal{T}}=\left(\alpha_1,\hdots, \alpha_L,\alpha^{\dagger}_1,\hdots, \alpha^{\dagger}_L\right)$, $K$ is a $2L\times 2L$ matrix, and
$$
\sigma = \left(\begin{array}{cc}
0_{L\times L} & \mathbbm{1}_{L\times L} \\ 
\mathbbm{1}_{L\times L} & 0_{L\times L}
\end{array}  \right).
$$
In order to bring \eqref{equ:twoptF} into the form of \eqref{equ:Blaizot}, we write the Hamiltonian as:
\begin{equation}
\label{equ:nondiagH}
H=\omega_s \beta^{\dagger}_s \beta_s = \omega_s O_{sq}^{*}O_{sp}\alpha^{\dagger}_{q} \alpha_{p} \equiv H_{qp} \alpha^{\dagger}_{q} \alpha_{p},
\end{equation}
where in the last step we have defined $H_{qp} \equiv \omega_s O_{sq}^{*}O_{sp}$.
It is convenient to choose $K$ such that $\left( K^{*}\right)^{\text{T}}=\bar{K}^{*} = -K$. To achieve this, we symmetrise the Hamiltonian, making use of the anticommutation properties of the fermionic operators:
\begin{equation}
e^{iHt}=e^{iH_{qp}\alpha^{\dagger}_{q}\alpha_{p}t}=e^{iH_{qp}\left(\frac{1}{2}\alpha^{\dagger}_{q}\alpha_{p}-\frac{1}{2}\alpha_{p}\alpha_{q}^{\dagger}+\frac{1}{2}\delta_{qp}\right)t}.
\end{equation}
In this form we have $e^{\frac{1}{2}\tilde{\nu}K\nu}$ with $K=\left(\begin{array}{cc}
0 & -i\bar{H}t \\ 
iHt & 0
\end{array}\right) $ and hence obtain: $e^{-\sigma K}=\left(\begin{array}{cc}
e^{-iHt} & 0 \\ 
0 & e^{i\bar{H}t}
\end{array}  \right).$

We now are in a position to apply \eqref{equ:Blaizot} to \eqref{equ:twoptF}, which leaves us with the following equation:
\begin{equation}
C_{ij}(t) = A_{kiqj} \sum\limits^{2L}_{m,n}\left(e^{-\sigma K}\right)_{k+L,m}\left(e^{-\sigma K} \right)_{q,n} Q_{mn},
\end{equation}
where the matrix element $Q_{mn}$ is defined as follows:
\begin{equation}
 Q_{mn} =\langle \psi_0^{(N)}\vert \nu_{m}\nu_{n}\vert \psi_0^{(N)}\rangle,
\end{equation}
and $A_{kiqj} =\phi_{ki}\phi^{*}_{qj}$.
The sums are restricted due to the shape of $K$ and the Fermi-energy, limiting when $Q_{mn}$ is non-zero.
We therefore obtain:
\begin{equation}
C_{ij}(t) = \sum\limits_{m=1}^{N}\sum\limits_{k=1}^{L}\sum\limits_{q=1}^{L}\phi_{ki}\phi^{*}_{qj}\left(e^{i\bar{H}}\right)_{km}\left(e^{-iH}\right)_{qm}
\end{equation}
As a final step we diagonalize the Hamiltonian by reversing \eqref{equ:nondiagH} using \eqref{equ:transformations} which allows us to transform the $\phi$'s and write the solution in the following form:
\begin{equation}
\label{equ:correl}
C_{ij}(t) = \sum\limits_{a,c=1}^{L}\sum\limits_{m=1}^{N}O_{am}
O_{cm}^{*}e^{-i(\omega_{a}-\omega_{c})t}\psi_{ci}\psi_{aj}^{*}.
\end{equation}
Setting $i=j$ in this formula recovers the expression for the density $\rho_j(t)$ given in (\ref{equ:density}).

\section{A toy calculation of the effect of disorder on the jump efficiency for small jump sizes}
\label{app:jumpeff}
\noindent
In this final appendix, we present a toy calculation that allows us to understand the $1 - \alpha W^2$ dependence of the jump efficiency at small jump sizes.

It represents the pre-quench single-particle eigenfunctions by:
\begin{eqnarray}
\phi_{nj}^{{+}} & = & \frac{1}{\sqrt{2}}\left(\delta_{j,j_0+n}+\delta_{j,j_0-n}\right);  \\
\phi_{nj}^{{-}} & = & \frac{1}{\sqrt{2}}\left(\delta_{j,j_0+n}-\delta_{j,j_0-n}\right).
\end{eqnarray}
We have denoted the symmetric and antisymmetric eigenfunctions separately, while the quantum number $n = 1,2,3,...$ (we ignore the $n=0$ case).
Essentially, this amounts to a cartoon of each harmonic oscillator eigenfunction in the form of two peaks at its classical turning points, retaining the information about whether the function is symmetric or antisymmetric.
We thus obtain the densities:
\begin{equation}
\vert\phi_{nj}^{{+}}\vert^2 = \vert\phi_{nj}^{{-}}\vert^2 = \frac{1}{2}\left(\delta_{j,j_0+n}+\delta_{j,j_0-n}\right).
\end{equation}
We need to occupy the symmetric and antisymmetric versions of $N/2$ eigenstates to get the correct particle number, i.e.\ we occupy the states with $1 \leqslant n \leqslant N/2$, such that the total density becomes
\begin{equation}
\rho_j = \sum_{n=1}^{N/2}\left(\delta_{j,j_0-n}+\delta_{j,j_0+n}\right).
\end{equation}
This is a `block' in real space, covering the region $j_0 - N/2 \leqslant j \leqslant j_0 + N/2$ (with a hole at $j = j_0$, but this causes only a $1/N$ effect, which we neglect).

Second, we assume that the post-quench eigenfunctions in the presence of disorder may similarly be approximated by:
\begin{eqnarray}
\psi_{nj}^{{+}} & = & \sqrt{\frac{1 + \sigma_n W}{2}} \,\delta_{j,j_1+n}+\sqrt{\frac{1 - \sigma_n W}{2}} \,\delta_{j,j_1-n}; \nonumber \\
& & \\
\psi_{nj}^{{-}} & = & \sqrt{\frac{1 + \sigma_n W}{2}} \,\delta_{j,j_1+n}-\sqrt{\frac{1 - \sigma_n W}{2}} \,\delta_{j,j_1-n}, \nonumber \\
& & 
\end{eqnarray}
where the random variable $\sigma_n = \pm 1$ is chosen independently for each value of $n$ to encode the presence of disorder.
Note that for both the pre- and post-quench eigenfunctions we have made the simplifying assumption that the position of the classical turning points is proportional to the energy of the eigenstate.  This corresponds to choosing a linear trapping potential rather than a quadratic one.

With the above choice of eigenfunctions we can determine $x_{1,n}$, the disorder-dependent center of mass of post-quench eigenfunction $n$:
\begin{equation}
x_{1,n} = \sum_{j}j\vert \psi^{{+}}_{nj}\vert^2 = \sum_{j}j\vert \psi^{{-}}_{nj}\vert^2 = j_{1} + \sigma_n n W,
\end{equation}
which is linear in $W$ and independent of the symmetry of the eigenfunction.

In order to determine the post-quench occupation function $n_\beta$, we first determine the overlap between a particular pair of pre- and post-quench eigenfunctions:
\begin{eqnarray}
O_{n_0 n_1}^{S_0 S_1} & \equiv & \sum_{j}\phi_{n_0 j}^{S_0}\psi_{n_1 j}^{S_1} \\
& = & \frac{\sqrt{1 + \sigma_{n_1} W}}{2} \left( \delta_{j_0+n_0,j_1+n_1}+ S_{0}\delta_{j_0-n_0,j_1+n_1} \right) \nonumber \\
& & \!\!\!\!\!\! + \frac{\sqrt{1 - \sigma_{n_1} W}}{2} S_1 \left( \delta_{j_0+n_0,j_1-n_1}+S_{0}\delta_{j_0-n_0,j_1-n_1} \right). \nonumber \\
& & 
\end{eqnarray}
Here $S_0,S_1 \in \{ {-}1,{+}1 \}$ are the symmetries of the pre- and post-quench eigenfunctions, and $n_0$ and $n_1$ are their quantum numbers.
Assuming without loss of generality that $j_1 > j_{0}$, the pure $S_{0}$ term is  always zero, so that we obtain as squared overlap:
\begin{eqnarray}
\left\vert O_{n_0 n_1}^{S_0 S_1} \right\vert^2 & = & \frac{1 + \sigma_{n_1} W}{4}\delta_{j_0+n_0,j_1+n_1} \nonumber \\
& & {+}\,\frac{1 - \sigma_{n_1} W}{4} \left( \delta_{j_0+n_0,j_1-n_1}+\delta_{j_0-n_0,j_1-n_1} \right). \nonumber \\
& & 
\end{eqnarray}
The post-quench occupation function then takes the form
\begin{eqnarray}
n_{n_1}^{S_1} & = & \sum_{n_0 = 1}^{N/2}\sum_{S_{0}=\pm1} \left( \frac{1 + \sigma_{n_1} W}{4}\delta_{j_0+n_0,j_1+n_1} \right. \nonumber \\
& & \left. {+}\,\frac{1 - \sigma_{n_1} W}{4} \left( \delta_{j_0+n_0,j_1-n_1}+\delta_{j_0-n_0,j_1-n_1} \right) \right). \nonumber \\
& & 
\end{eqnarray}
Since $n_1$ has to be positive, we obtain
\begin{equation}
n_{n_1}^{S_1} = \begin{cases}
\displaystyle   1 & n_1\leq N/2-\Delta; \\
& \\
\displaystyle    \frac{1 - \sigma_{n_1} W}{2}~ & N/2-\Delta < n_1 \leq N/2 + \Delta; \\
& \\
\displaystyle    0~ & \text{otherwise}.
  \end{cases}
\end{equation}
Hence the post-quench centre of mass is
\begin{eqnarray}
x_1 & = & \frac{2}{N} \sum_{n_1 = 1}^{L/2}\left(j_1 + \sigma_{n_1} n_1 W \right)n_{n_1}^{s_1} \\
& = & \frac{2}{N} \sum_{n_1 = 1}^{N/2-\Delta}\left(j_1 + \sigma_{n_1} n_1 W \right) \nonumber \\
& & \!\!\!\!\!\! {+}\,\frac{2}{N} \sum_{n_1 = N/2-\Delta+1}^{N/2+\Delta}\left(j_1 + \sigma_{n_1} n_1 W \right)\left(\frac{1 - \sigma_{n_1} W}{2}\right). \nonumber \\
& & 
\end{eqnarray}
Upon disorder-averaging, any term containing an odd power of $\sigma_{n_1}$ vanishes, while the average of any even power of $\sigma_{n_1}$ is unity.  Hence
\begin{eqnarray}
\overline{x_1\mathstrut} & = & \frac{1}{N} \left[ 2 j_1 \left(\frac{N}{2}-\Delta\right)+j_1\left(2\Delta\right)\right] \nonumber \\
& & \qquad \qquad \qquad {-}\,\frac{W^2}{N}\sum_{n_1 = N/2-\Delta+1}^{N/2+\Delta} n_1 \\
& = & j_1 - \Delta W^2 \frac{N+1}{N} \\
& \approx & j_1 - \Delta W^2.
\end{eqnarray}

The jump efficiency is given by the difference between this post-quench center of mass and the pre-quench one in units of the jump size:
\bea
\eta & \equiv & \frac{\overline{x_1\mathstrut}-j_0}{\Delta} \\
& \approx & \frac{\Delta - \Delta W^2}{\Delta} \\
& = & 1 - W^2.
\eea
This is the qualitative behavior that we observe for small jump sizes in Fig.~\ref{fig:jump_eff}.
\bibliography{2014_quench_noint_v3}{}

\begin{thebibliography}{50}%
\makeatletter
\providecommand \@ifxundefined [1]{%
 \@ifx{#1\undefined}
}%
\providecommand \@ifnum [1]{%
 \ifnum #1\expandafter \@firstoftwo
 \else \expandafter \@secondoftwo
 \fi
}%
\providecommand \@ifx [1]{%
 \ifx #1\expandafter \@firstoftwo
 \else \expandafter \@secondoftwo
 \fi
}%
\providecommand \natexlab [1]{#1}%
\providecommand \enquote  [1]{``#1''}%
\providecommand \bibnamefont  [1]{#1}%
\providecommand \bibfnamefont [1]{#1}%
\providecommand \citenamefont [1]{#1}%
\providecommand \href@noop [0]{\@secondoftwo}%
\providecommand \href [0]{\begingroup \@sanitize@url \@href}%
\providecommand \@href[1]{\@@startlink{#1}\@@href}%
\providecommand \@@href[1]{\endgroup#1\@@endlink}%
\providecommand \@sanitize@url [0]{\catcode `\\12\catcode `\$12\catcode
  `\&12\catcode `\#12\catcode `\^12\catcode `\_12\catcode `\%12\relax}%
\providecommand \@@startlink[1]{}%
\providecommand \@@endlink[0]{}%
\providecommand \url  [0]{\begingroup\@sanitize@url \@url }%
\providecommand \@url [1]{\endgroup\@href {#1}{\urlprefix }}%
\providecommand \urlprefix  [0]{URL }%
\providecommand \Eprint [0]{\href }%
\providecommand \doibase [0]{http://dx.doi.org/}%
\providecommand \selectlanguage [0]{\@gobble}%
\providecommand \bibinfo  [0]{\@secondoftwo}%
\providecommand \bibfield  [0]{\@secondoftwo}%
\providecommand \translation [1]{[#1]}%
\providecommand \BibitemOpen [0]{}%
\providecommand \bibitemStop [0]{}%
\providecommand \bibitemNoStop [0]{.\EOS\space}%
\providecommand \EOS [0]{\spacefactor3000\relax}%
\providecommand \BibitemShut  [1]{\csname bibitem#1\endcsname}%
\let\auto@bib@innerbib\@empty
\bibitem [{\citenamefont {Gornyi}\ \emph {et~al.}(2005)\citenamefont {Gornyi},
  \citenamefont {Mirlin},\ and\ \citenamefont {Polyakov}}]{Gornyi2005}%
  \BibitemOpen
  \bibfield  {author} {\bibinfo {author} {\bibfnamefont {I.~V.}\ \bibnamefont
  {Gornyi}}, \bibinfo {author} {\bibfnamefont {A.~D.}\ \bibnamefont {Mirlin}},
  \ and\ \bibinfo {author} {\bibfnamefont {D.~G.}\ \bibnamefont {Polyakov}},\
  }\href {\doibase 10.1103/PhysRevLett.95.206603} {\bibfield  {journal}
  {\bibinfo  {journal} {Phys. Rev. Lett.}\ }\textbf {\bibinfo {volume} {95}},\
  \bibinfo {pages} {206603} (\bibinfo {year} {2005})}\BibitemShut {NoStop}%
\bibitem [{\citenamefont {Basko}\ \emph {et~al.}(2006)\citenamefont {Basko},
  \citenamefont {Aleiner},\ and\ \citenamefont {Altshuler}}]{Basko2006}%
  \BibitemOpen
  \bibfield  {author} {\bibinfo {author} {\bibfnamefont {D.~M.}\ \bibnamefont
  {Basko}}, \bibinfo {author} {\bibfnamefont {I.~L.}\ \bibnamefont {Aleiner}},
  \ and\ \bibinfo {author} {\bibfnamefont {B.~L.}\ \bibnamefont {Altshuler}},\
  }\href {\doibase 10.1016/j.aop.2005.11.014} {\bibfield  {journal} {\bibinfo
  {journal} {Ann. Phys. (N. Y).}\ }\textbf {\bibinfo {volume} {321}},\ \bibinfo
  {pages} {1126} (\bibinfo {year} {2006})}\BibitemShut {NoStop}%
\bibitem [{\citenamefont {Rigol}\ \emph {et~al.}(2008)\citenamefont {Rigol},
  \citenamefont {Dunjko},\ and\ \citenamefont {Olshanii}}]{Rigol2008}%
  \BibitemOpen
  \bibfield  {author} {\bibinfo {author} {\bibfnamefont {M.}~\bibnamefont
  {Rigol}}, \bibinfo {author} {\bibfnamefont {V.}~\bibnamefont {Dunjko}}, \
  and\ \bibinfo {author} {\bibfnamefont {M.}~\bibnamefont {Olshanii}},\ }\href
  {\doibase 10.1038/nature06838} {\bibfield  {journal} {\bibinfo  {journal}
  {Nature}\ }\textbf {\bibinfo {volume} {452}},\ \bibinfo {pages} {854}
  (\bibinfo {year} {2008})}\BibitemShut {NoStop}%
\bibitem [{\citenamefont {Polkovnikov}\ \emph {et~al.}(2011)\citenamefont
  {Polkovnikov}, \citenamefont {Sengupta}, \citenamefont {Silva},\ and\
  \citenamefont {Vengalattore}}]{Polkovnikov2011}%
  \BibitemOpen
  \bibfield  {author} {\bibinfo {author} {\bibfnamefont {A.}~\bibnamefont
  {Polkovnikov}}, \bibinfo {author} {\bibfnamefont {K.}~\bibnamefont
  {Sengupta}}, \bibinfo {author} {\bibfnamefont {A.}~\bibnamefont {Silva}}, \
  and\ \bibinfo {author} {\bibfnamefont {M.}~\bibnamefont {Vengalattore}},\
  }\href {\doibase 10.1103/RevModPhys.83.863} {\bibfield  {journal} {\bibinfo
  {journal} {Rev. Mod. Phys.}\ }\textbf {\bibinfo {volume} {83}},\ \bibinfo
  {pages} {863} (\bibinfo {year} {2011})}\BibitemShut {NoStop}%
\bibitem [{\citenamefont {Huse}\ \emph {et~al.}(2014)\citenamefont {Huse},
  \citenamefont {Nandkishore},\ and\ \citenamefont {Oganesyan}}]{Huse2014a}%
  \BibitemOpen
  \bibfield  {author} {\bibinfo {author} {\bibfnamefont {D.~A.}\ \bibnamefont
  {Huse}}, \bibinfo {author} {\bibfnamefont {R.}~\bibnamefont {Nandkishore}}, \
  and\ \bibinfo {author} {\bibfnamefont {V.}~\bibnamefont {Oganesyan}},\ }\href
  {\doibase 10.1103/PhysRevB.90.174202} {\bibfield  {journal} {\bibinfo
  {journal} {Phys. Rev. B}\ }\textbf {\bibinfo {volume} {90}},\ \bibinfo
  {pages} {174202} (\bibinfo {year} {2014})}\BibitemShut {NoStop}%
\bibitem [{\citenamefont {Nandkishore}\ and\ \citenamefont
  {Huse}(2015)}]{Nandkishore2015}%
  \BibitemOpen
  \bibfield  {author} {\bibinfo {author} {\bibfnamefont {R.}~\bibnamefont
  {Nandkishore}}\ and\ \bibinfo {author} {\bibfnamefont {D.~A.}\ \bibnamefont
  {Huse}},\ }\href {\doibase 10.1146/annurev-conmatphys-031214-014726}
  {\bibfield  {journal} {\bibinfo  {journal} {Annu. Rev. Condens. Matter
  Phys.}\ }\textbf {\bibinfo {volume} {6}},\ \bibinfo {pages} {15} (\bibinfo
  {year} {2015})}\BibitemShut {NoStop}%
\bibitem [{\citenamefont {Eisert}\ \emph {et~al.}(2015)\citenamefont {Eisert},
  \citenamefont {Friesdorf},\ and\ \citenamefont {Gogolin}}]{Eisert2015}%
  \BibitemOpen
  \bibfield  {author} {\bibinfo {author} {\bibfnamefont {J.}~\bibnamefont
  {Eisert}}, \bibinfo {author} {\bibfnamefont {M.}~\bibnamefont {Friesdorf}}, \
  and\ \bibinfo {author} {\bibfnamefont {C.}~\bibnamefont {Gogolin}},\ }\href
  {\doibase 10.1038/nphys3215} {\bibfield  {journal} {\bibinfo  {journal} {Nat.
  Phys.}\ }\textbf {\bibinfo {volume} {11}},\ \bibinfo {pages} {124} (\bibinfo
  {year} {2015})}\BibitemShut {NoStop}%
\bibitem [{\citenamefont {Kinoshita}\ \emph {et~al.}(2006)\citenamefont
  {Kinoshita}, \citenamefont {Wenger},\ and\ \citenamefont
  {Weiss}}]{Kinoshita2006}%
  \BibitemOpen
  \bibfield  {author} {\bibinfo {author} {\bibfnamefont {T.}~\bibnamefont
  {Kinoshita}}, \bibinfo {author} {\bibfnamefont {T.}~\bibnamefont {Wenger}}, \
  and\ \bibinfo {author} {\bibfnamefont {D.~S.}\ \bibnamefont {Weiss}},\ }\href
  {\doibase 10.1038/nature04693} {\bibfield  {journal} {\bibinfo  {journal}
  {Nature}\ }\textbf {\bibinfo {volume} {440}},\ \bibinfo {pages} {900}
  (\bibinfo {year} {2006})}\BibitemShut {NoStop}%
\bibitem [{\citenamefont {Billy}\ \emph {et~al.}(2008)\citenamefont {Billy},
  \citenamefont {Josse}, \citenamefont {Zuo}, \citenamefont {Bernard},
  \citenamefont {Hambrecht}, \citenamefont {Lugan}, \citenamefont
  {Cl{\'{e}}ment}, \citenamefont {Sanchez-Palencia}, \citenamefont {Bouyer},\
  and\ \citenamefont {Aspect}}]{Billy2008}%
  \BibitemOpen
  \bibfield  {author} {\bibinfo {author} {\bibfnamefont {J.}~\bibnamefont
  {Billy}}, \bibinfo {author} {\bibfnamefont {V.}~\bibnamefont {Josse}},
  \bibinfo {author} {\bibfnamefont {Z.}~\bibnamefont {Zuo}}, \bibinfo {author}
  {\bibfnamefont {A.}~\bibnamefont {Bernard}}, \bibinfo {author} {\bibfnamefont
  {B.}~\bibnamefont {Hambrecht}}, \bibinfo {author} {\bibfnamefont
  {P.}~\bibnamefont {Lugan}}, \bibinfo {author} {\bibfnamefont
  {D.}~\bibnamefont {Cl{\'{e}}ment}}, \bibinfo {author} {\bibfnamefont
  {L.}~\bibnamefont {Sanchez-Palencia}}, \bibinfo {author} {\bibfnamefont
  {P.}~\bibnamefont {Bouyer}}, \ and\ \bibinfo {author} {\bibfnamefont
  {A.}~\bibnamefont {Aspect}},\ }\href {\doibase 10.1038/nature07000}
  {\bibfield  {journal} {\bibinfo  {journal} {Nature}\ }\textbf {\bibinfo
  {volume} {453}},\ \bibinfo {pages} {891} (\bibinfo {year}
  {2008})}\BibitemShut {NoStop}%
\bibitem [{\citenamefont {Roati}\ \emph {et~al.}(2008)\citenamefont {Roati},
  \citenamefont {D'Errico}, \citenamefont {Fallani}, \citenamefont {Fattori},
  \citenamefont {Fort}, \citenamefont {Zaccanti}, \citenamefont {Modugno},
  \citenamefont {Modugno},\ and\ \citenamefont {Inguscio}}]{Roati2008}%
  \BibitemOpen
  \bibfield  {author} {\bibinfo {author} {\bibfnamefont {G.}~\bibnamefont
  {Roati}}, \bibinfo {author} {\bibfnamefont {C.}~\bibnamefont {D'Errico}},
  \bibinfo {author} {\bibfnamefont {L.}~\bibnamefont {Fallani}}, \bibinfo
  {author} {\bibfnamefont {M.}~\bibnamefont {Fattori}}, \bibinfo {author}
  {\bibfnamefont {C.}~\bibnamefont {Fort}}, \bibinfo {author} {\bibfnamefont
  {M.}~\bibnamefont {Zaccanti}}, \bibinfo {author} {\bibfnamefont
  {G.}~\bibnamefont {Modugno}}, \bibinfo {author} {\bibfnamefont
  {M.}~\bibnamefont {Modugno}}, \ and\ \bibinfo {author} {\bibfnamefont
  {M.}~\bibnamefont {Inguscio}},\ }\href {\doibase 10.1038/nature07071}
  {\bibfield  {journal} {\bibinfo  {journal} {Nature}\ }\textbf {\bibinfo
  {volume} {453}},\ \bibinfo {pages} {895} (\bibinfo {year}
  {2008})}\BibitemShut {NoStop}%
\bibitem [{\citenamefont {Schreiber}\ \emph {et~al.}(2015)\citenamefont
  {Schreiber}, \citenamefont {Hodgman}, \citenamefont {Bordia}, \citenamefont
  {L{\"{u}}schen}, \citenamefont {Fischer}, \citenamefont {Vosk}, \citenamefont
  {Altman}, \citenamefont {Schneider},\ and\ \citenamefont
  {Bloch}}]{Schreiber2015}%
  \BibitemOpen
  \bibfield  {author} {\bibinfo {author} {\bibfnamefont {M.}~\bibnamefont
  {Schreiber}}, \bibinfo {author} {\bibfnamefont {S.~S.}\ \bibnamefont
  {Hodgman}}, \bibinfo {author} {\bibfnamefont {P.}~\bibnamefont {Bordia}},
  \bibinfo {author} {\bibfnamefont {H.~P.}\ \bibnamefont {L{\"{u}}schen}},
  \bibinfo {author} {\bibfnamefont {M.~H.}\ \bibnamefont {Fischer}}, \bibinfo
  {author} {\bibfnamefont {R.}~\bibnamefont {Vosk}}, \bibinfo {author}
  {\bibfnamefont {E.}~\bibnamefont {Altman}}, \bibinfo {author} {\bibfnamefont
  {U.}~\bibnamefont {Schneider}}, \ and\ \bibinfo {author} {\bibfnamefont
  {I.}~\bibnamefont {Bloch}},\ }\href {\doibase 10.1126/science.aaa7432}
  {\bibfield  {journal} {\bibinfo  {journal} {Science}\ }\textbf {\bibinfo
  {volume} {349}},\ \bibinfo {pages} {842} (\bibinfo {year}
  {2015})}\BibitemShut {NoStop}%
\bibitem [{\citenamefont {Kondov}\ \emph {et~al.}(2015)\citenamefont {Kondov},
  \citenamefont {McGehee}, \citenamefont {Xu},\ and\ \citenamefont
  {DeMarco}}]{Kondov2015}%
  \BibitemOpen
  \bibfield  {author} {\bibinfo {author} {\bibfnamefont {S.~S.}\ \bibnamefont
  {Kondov}}, \bibinfo {author} {\bibfnamefont {W.~R.}\ \bibnamefont {McGehee}},
  \bibinfo {author} {\bibfnamefont {W.}~\bibnamefont {Xu}}, \ and\ \bibinfo
  {author} {\bibfnamefont {B.}~\bibnamefont {DeMarco}},\ }\href {\doibase
  10.1103/PhysRevLett.114.083002} {\bibfield  {journal} {\bibinfo  {journal}
  {Phys. Rev. Lett.}\ }\textbf {\bibinfo {volume} {114}},\ \bibinfo {pages}
  {206603} (\bibinfo {year} {2015})}\BibitemShut {NoStop}%
\bibitem [{\citenamefont {Childress}\ \emph {et~al.}(2006)\citenamefont
  {Childress}, \citenamefont {{Gurudev Dutt}}, \citenamefont {Taylor},
  \citenamefont {Zibrov}, \citenamefont {Jelezko}, \citenamefont {Wrachtrup},
  \citenamefont {Hemmer},\ and\ \citenamefont {Lukin}}]{Childress2006}%
  \BibitemOpen
  \bibfield  {author} {\bibinfo {author} {\bibfnamefont {L.}~\bibnamefont
  {Childress}}, \bibinfo {author} {\bibfnamefont {M.~V.}\ \bibnamefont
  {{Gurudev Dutt}}}, \bibinfo {author} {\bibfnamefont {J.~M.}\ \bibnamefont
  {Taylor}}, \bibinfo {author} {\bibfnamefont {A.~S.}\ \bibnamefont {Zibrov}},
  \bibinfo {author} {\bibfnamefont {F.}~\bibnamefont {Jelezko}}, \bibinfo
  {author} {\bibfnamefont {J.}~\bibnamefont {Wrachtrup}}, \bibinfo {author}
  {\bibfnamefont {P.~R.}\ \bibnamefont {Hemmer}}, \ and\ \bibinfo {author}
  {\bibfnamefont {M.~D.}\ \bibnamefont {Lukin}},\ }\href {\doibase
  10.1126/science.1131871} {\bibfield  {journal} {\bibinfo  {journal}
  {Science}\ }\textbf {\bibinfo {volume} {314}},\ \bibinfo {pages} {281}
  (\bibinfo {year} {2006})}\BibitemShut {NoStop}%
\bibitem [{\citenamefont {Berry}(1977)}]{Berry1977}%
  \BibitemOpen
  \bibfield  {author} {\bibinfo {author} {\bibfnamefont {M.~V.}\ \bibnamefont
  {Berry}},\ }\href {\doibase 10.1088/0305-4470/10/12/016} {\bibfield
  {journal} {\bibinfo  {journal} {J. Phys. A}\ }\textbf {\bibinfo {volume}
  {10}},\ \bibinfo {pages} {2083} (\bibinfo {year} {1977})}\BibitemShut
  {NoStop}%
\bibitem [{\citenamefont {Deutsch}(1991)}]{Deutsch1991}%
  \BibitemOpen
  \bibfield  {author} {\bibinfo {author} {\bibfnamefont {J.~M.}\ \bibnamefont
  {Deutsch}},\ }\href {\doibase 10.1103/PhysRevA.43.2046} {\bibfield  {journal}
  {\bibinfo  {journal} {Phys. Rev. A}\ }\textbf {\bibinfo {volume} {43}},\
  \bibinfo {pages} {2046} (\bibinfo {year} {1991})}\BibitemShut {NoStop}%
\bibitem [{\citenamefont {Srednicki}(1994)}]{Srednicki1994}%
  \BibitemOpen
  \bibfield  {author} {\bibinfo {author} {\bibfnamefont {M.}~\bibnamefont
  {Srednicki}},\ }\href {\doibase 10.1103/PhysRevE.50.888} {\bibfield
  {journal} {\bibinfo  {journal} {Phys. Rev. E}\ }\textbf {\bibinfo {volume}
  {50}},\ \bibinfo {pages} {888} (\bibinfo {year} {1994})}\BibitemShut
  {NoStop}%
\bibitem [{\citenamefont {Pal}\ and\ \citenamefont {Huse}(2010)}]{Pal2010}%
  \BibitemOpen
  \bibfield  {author} {\bibinfo {author} {\bibfnamefont {A.}~\bibnamefont
  {Pal}}\ and\ \bibinfo {author} {\bibfnamefont {D.~A.}\ \bibnamefont {Huse}},\
  }\href {\doibase 10.1103/PhysRevB.82.174411} {\bibfield  {journal} {\bibinfo
  {journal} {Phys. Rev. B}\ }\textbf {\bibinfo {volume} {82}},\ \bibinfo
  {pages} {174411} (\bibinfo {year} {2010})}\BibitemShut {NoStop}%
\bibitem [{\citenamefont {Bardarson}\ \emph {et~al.}(2012)\citenamefont
  {Bardarson}, \citenamefont {Pollmann},\ and\ \citenamefont
  {Moore}}]{Bardarson2012}%
  \BibitemOpen
  \bibfield  {author} {\bibinfo {author} {\bibfnamefont {J.~H.}\ \bibnamefont
  {Bardarson}}, \bibinfo {author} {\bibfnamefont {F.}~\bibnamefont {Pollmann}},
  \ and\ \bibinfo {author} {\bibfnamefont {J.~E.}\ \bibnamefont {Moore}},\
  }\href {\doibase 10.1103/PhysRevLett.109.017202} {\bibfield  {journal}
  {\bibinfo  {journal} {Phys. Rev. Lett.}\ }\textbf {\bibinfo {volume} {109}},\
  \bibinfo {pages} {017202} (\bibinfo {year} {2012})}\BibitemShut {NoStop}%
\bibitem [{\citenamefont {Vosk}\ and\ \citenamefont {Altman}(2013)}]{Vosk2013}%
  \BibitemOpen
  \bibfield  {author} {\bibinfo {author} {\bibfnamefont {R.}~\bibnamefont
  {Vosk}}\ and\ \bibinfo {author} {\bibfnamefont {E.}~\bibnamefont {Altman}},\
  }\href {\doibase 10.1103/PhysRevLett.110.067204} {\bibfield  {journal}
  {\bibinfo  {journal} {Phys. Rev. Lett.}\ }\textbf {\bibinfo {volume} {110}},\
  \bibinfo {pages} {067204} (\bibinfo {year} {2013})}\BibitemShut {NoStop}%
\bibitem [{\citenamefont {Chandran}\ \emph {et~al.}(2014)\citenamefont
  {Chandran}, \citenamefont {Khemani}, \citenamefont {Laumann},\ and\
  \citenamefont {Sondhi}}]{Chandran2014}%
  \BibitemOpen
  \bibfield  {author} {\bibinfo {author} {\bibfnamefont {A.}~\bibnamefont
  {Chandran}}, \bibinfo {author} {\bibfnamefont {V.}~\bibnamefont {Khemani}},
  \bibinfo {author} {\bibfnamefont {C.~R.}\ \bibnamefont {Laumann}}, \ and\
  \bibinfo {author} {\bibfnamefont {S.~L.}\ \bibnamefont {Sondhi}},\ }\href
  {\doibase 10.1103/PhysRevB.89.144201} {\bibfield  {journal} {\bibinfo
  {journal} {Phys. Rev. B}\ }\textbf {\bibinfo {volume} {89}},\ \bibinfo
  {pages} {144201} (\bibinfo {year} {2014})}\BibitemShut {NoStop}%
\bibitem [{\citenamefont {Ponte}\ \emph {et~al.}(2015)\citenamefont {Ponte},
  \citenamefont {Papi{\'{c}}}, \citenamefont {Huveneers},\ and\ \citenamefont
  {Abanin}}]{Ponte2015}%
  \BibitemOpen
  \bibfield  {author} {\bibinfo {author} {\bibfnamefont {P.}~\bibnamefont
  {Ponte}}, \bibinfo {author} {\bibfnamefont {Z.}~\bibnamefont {Papi{\'{c}}}},
  \bibinfo {author} {\bibfnamefont {F.}~\bibnamefont {Huveneers}}, \ and\
  \bibinfo {author} {\bibfnamefont {D.~A.}\ \bibnamefont {Abanin}},\ }\href
  {\doibase 10.1103/PhysRevLett.114.140401} {\bibfield  {journal} {\bibinfo
  {journal} {Phys. Rev. Lett.}\ }\textbf {\bibinfo {volume} {114}},\ \bibinfo
  {pages} {140401} (\bibinfo {year} {2015})}\BibitemShut {NoStop}%
\bibitem [{\citenamefont {Agarwal}\ \emph {et~al.}(2015)\citenamefont
  {Agarwal}, \citenamefont {Gopalakrishnan}, \citenamefont {Knap},
  \citenamefont {M{\"{u}}ller},\ and\ \citenamefont {Demler}}]{Agarwal2015}%
  \BibitemOpen
  \bibfield  {author} {\bibinfo {author} {\bibfnamefont {K.}~\bibnamefont
  {Agarwal}}, \bibinfo {author} {\bibfnamefont {S.}~\bibnamefont
  {Gopalakrishnan}}, \bibinfo {author} {\bibfnamefont {M.}~\bibnamefont
  {Knap}}, \bibinfo {author} {\bibfnamefont {M.}~\bibnamefont {M{\"{u}}ller}},
  \ and\ \bibinfo {author} {\bibfnamefont {E.}~\bibnamefont {Demler}},\ }\href
  {\doibase 10.1103/PhysRevLett.114.160401} {\bibfield  {journal} {\bibinfo
  {journal} {Phys. Rev. Lett.}\ }\textbf {\bibinfo {volume} {114}},\ \bibinfo
  {pages} {160401} (\bibinfo {year} {2015})}\BibitemShut {NoStop}%
\bibitem [{\citenamefont {Imbrie}(2016)}]{Imbrie2016}%
  \BibitemOpen
  \bibfield  {author} {\bibinfo {author} {\bibfnamefont {J.~Z.}\ \bibnamefont
  {Imbrie}},\ }\href {\doibase 10.1103/PhysRevLett.110.067204} {\bibfield
  {journal} {\bibinfo  {journal} {J. Stat. Phys.}\ }\textbf {\bibinfo {volume}
  {163}},\ \bibinfo {pages} {998} (\bibinfo {year} {2016})}\BibitemShut
  {NoStop}%
\bibitem [{\citenamefont {Rigol}\ \emph {et~al.}(2007)\citenamefont {Rigol},
  \citenamefont {Dunjko}, \citenamefont {Yurovsky},\ and\ \citenamefont
  {Olshanii}}]{Rigol2007}%
  \BibitemOpen
  \bibfield  {author} {\bibinfo {author} {\bibfnamefont {M.}~\bibnamefont
  {Rigol}}, \bibinfo {author} {\bibfnamefont {V.}~\bibnamefont {Dunjko}},
  \bibinfo {author} {\bibfnamefont {V.}~\bibnamefont {Yurovsky}}, \ and\
  \bibinfo {author} {\bibfnamefont {M.}~\bibnamefont {Olshanii}},\ }\href
  {\doibase 10.1103/PhysRevLett.98.050405} {\bibfield  {journal} {\bibinfo
  {journal} {Phys. Rev. Lett.}\ }\textbf {\bibinfo {volume} {98}},\ \bibinfo
  {pages} {050405} (\bibinfo {year} {2007})}\BibitemShut {NoStop}%
\bibitem [{\citenamefont {Caneva}\ \emph {et~al.}(2011)\citenamefont {Caneva},
  \citenamefont {Canovi}, \citenamefont {Rossini}, \citenamefont {Santoro},\
  and\ \citenamefont {Silva}}]{Caneva2011}%
  \BibitemOpen
  \bibfield  {author} {\bibinfo {author} {\bibfnamefont {T.}~\bibnamefont
  {Caneva}}, \bibinfo {author} {\bibfnamefont {E.}~\bibnamefont {Canovi}},
  \bibinfo {author} {\bibfnamefont {D.}~\bibnamefont {Rossini}}, \bibinfo
  {author} {\bibfnamefont {G.~E.}\ \bibnamefont {Santoro}}, \ and\ \bibinfo
  {author} {\bibfnamefont {A.}~\bibnamefont {Silva}},\ }\href {\doibase
  10.1088/1742-5468/2011/07/P07015} {\bibfield  {journal} {\bibinfo  {journal}
  {J. Stat. Mech. Theory Exp.}\ }\textbf {\bibinfo {volume} {2011}},\ \bibinfo
  {pages} {P07015} (\bibinfo {year} {2011})}\BibitemShut {NoStop}%
\bibitem [{\citenamefont {Cassidy}\ \emph {et~al.}(2011)\citenamefont
  {Cassidy}, \citenamefont {Clark},\ and\ \citenamefont {Rigol}}]{Cassidy2011}%
  \BibitemOpen
  \bibfield  {author} {\bibinfo {author} {\bibfnamefont {A.~C.}\ \bibnamefont
  {Cassidy}}, \bibinfo {author} {\bibfnamefont {C.~W.}\ \bibnamefont {Clark}},
  \ and\ \bibinfo {author} {\bibfnamefont {M.}~\bibnamefont {Rigol}},\ }\href
  {\doibase 10.1103/PhysRevLett.106.140405} {\bibfield  {journal} {\bibinfo
  {journal} {Phys. Rev. Lett.}\ }\textbf {\bibinfo {volume} {106}},\ \bibinfo
  {pages} {140405} (\bibinfo {year} {2011})}\BibitemShut {NoStop}%
\bibitem [{\citenamefont {Caux}\ and\ \citenamefont {Konik}(2012)}]{Caux2012}%
  \BibitemOpen
  \bibfield  {author} {\bibinfo {author} {\bibfnamefont {J.~S.}\ \bibnamefont
  {Caux}}\ and\ \bibinfo {author} {\bibfnamefont {R.~M.}\ \bibnamefont
  {Konik}},\ }\href {\doibase 10.1103/PhysRevLett.109.175301} {\bibfield
  {journal} {\bibinfo  {journal} {Phys. Rev. Lett.}\ }\textbf {\bibinfo
  {volume} {109}},\ \bibinfo {pages} {175301} (\bibinfo {year}
  {2012})}\BibitemShut {NoStop}%
\bibitem [{\citenamefont {Caux}\ and\ \citenamefont {Essler}(2013)}]{Caux2013}%
  \BibitemOpen
  \bibfield  {author} {\bibinfo {author} {\bibfnamefont {J.~S.}\ \bibnamefont
  {Caux}}\ and\ \bibinfo {author} {\bibfnamefont {F.~H.~L.}\ \bibnamefont
  {Essler}},\ }\href {\doibase 10.1103/PhysRevLett.110.257203} {\bibfield
  {journal} {\bibinfo  {journal} {Phys. Rev. Lett.}\ }\textbf {\bibinfo
  {volume} {110}},\ \bibinfo {pages} {257203} (\bibinfo {year}
  {2013})}\BibitemShut {NoStop}%
\bibitem [{\citenamefont {Pozsgay}\ \emph {et~al.}(2014)\citenamefont
  {Pozsgay}, \citenamefont {Mesty{\'{a}}n}, \citenamefont {Werner},
  \citenamefont {Kormos}, \citenamefont {Zar{\'{a}}nd},\ and\ \citenamefont
  {Tak{\'{a}}cs}}]{Pozsgay2014}%
  \BibitemOpen
  \bibfield  {author} {\bibinfo {author} {\bibfnamefont {B.}~\bibnamefont
  {Pozsgay}}, \bibinfo {author} {\bibfnamefont {M.}~\bibnamefont
  {Mesty{\'{a}}n}}, \bibinfo {author} {\bibfnamefont {M.~A.}\ \bibnamefont
  {Werner}}, \bibinfo {author} {\bibfnamefont {M.}~\bibnamefont {Kormos}},
  \bibinfo {author} {\bibfnamefont {G.}~\bibnamefont {Zar{\'{a}}nd}}, \ and\
  \bibinfo {author} {\bibfnamefont {G.}~\bibnamefont {Tak{\'{a}}cs}},\ }\href
  {\doibase 10.1103/PhysRevLett.113.117203} {\bibfield  {journal} {\bibinfo
  {journal} {Phys. Rev. Lett.}\ }\textbf {\bibinfo {volume} {113}},\ \bibinfo
  {pages} {117203} (\bibinfo {year} {2014})}\BibitemShut {NoStop}%
\bibitem [{\citenamefont {Essler}\ \emph {et~al.}(2015)\citenamefont {Essler},
  \citenamefont {Mussardo},\ and\ \citenamefont {Panfil}}]{Essler2015}%
  \BibitemOpen
  \bibfield  {author} {\bibinfo {author} {\bibfnamefont {F.~H.~L.}\
  \bibnamefont {Essler}}, \bibinfo {author} {\bibfnamefont {G.}~\bibnamefont
  {Mussardo}}, \ and\ \bibinfo {author} {\bibfnamefont {M.}~\bibnamefont
  {Panfil}},\ }\href {\doibase 10.1103/PhysRevA.91.051602} {\bibfield
  {journal} {\bibinfo  {journal} {Phys. Rev. A}\ }\textbf {\bibinfo {volume}
  {91}},\ \bibinfo {pages} {051602(R)} (\bibinfo {year} {2015})}\BibitemShut
  {NoStop}%
\bibitem [{\citenamefont {Lye}\ \emph {et~al.}(2005)\citenamefont {Lye},
  \citenamefont {Fallani}, \citenamefont {Modugno}, \citenamefont {Wiersma},
  \citenamefont {Fort},\ and\ \citenamefont {Inguscio}}]{Lye2005}%
  \BibitemOpen
  \bibfield  {author} {\bibinfo {author} {\bibfnamefont {J.~E.}\ \bibnamefont
  {Lye}}, \bibinfo {author} {\bibfnamefont {L.}~\bibnamefont {Fallani}},
  \bibinfo {author} {\bibfnamefont {M.}~\bibnamefont {Modugno}}, \bibinfo
  {author} {\bibfnamefont {D.~S.}\ \bibnamefont {Wiersma}}, \bibinfo {author}
  {\bibfnamefont {C.}~\bibnamefont {Fort}}, \ and\ \bibinfo {author}
  {\bibfnamefont {M.}~\bibnamefont {Inguscio}},\ }\href {\doibase
  10.1103/PhysRevLett.95.070401} {\bibfield  {journal} {\bibinfo  {journal}
  {Phys. Rev. Lett.}\ }\textbf {\bibinfo {volume} {95}},\ \bibinfo {pages}
  {070401} (\bibinfo {year} {2005})}\BibitemShut {NoStop}%
\bibitem [{\citenamefont {{Hecker Denschlag}}\ \emph
  {et~al.}(2002)\citenamefont {{Hecker Denschlag}}, \citenamefont {Simsarian},
  \citenamefont {H{\"{a}}ffner}, \citenamefont {McKenzi}, \citenamefont
  {Browaeys}, \citenamefont {Cho}, \citenamefont {Helmerson}, \citenamefont
  {Rolston},\ and\ \citenamefont {Phillips}}]{HeckerDenschlag2002}%
  \BibitemOpen
  \bibfield  {author} {\bibinfo {author} {\bibfnamefont {J.}~\bibnamefont
  {{Hecker Denschlag}}}, \bibinfo {author} {\bibfnamefont {J.~E.}\ \bibnamefont
  {Simsarian}}, \bibinfo {author} {\bibfnamefont {H.}~\bibnamefont
  {H{\"{a}}ffner}}, \bibinfo {author} {\bibfnamefont {C.}~\bibnamefont
  {McKenzi}}, \bibinfo {author} {\bibfnamefont {A.}~\bibnamefont {Browaeys}},
  \bibinfo {author} {\bibfnamefont {D.}~\bibnamefont {Cho}}, \bibinfo {author}
  {\bibfnamefont {K.}~\bibnamefont {Helmerson}}, \bibinfo {author}
  {\bibfnamefont {S.~L.}\ \bibnamefont {Rolston}}, \ and\ \bibinfo {author}
  {\bibfnamefont {W.~D.}\ \bibnamefont {Phillips}},\ }\href
  {http://iopscience.iop.org/0953-4075/35/14/307} {\bibfield  {journal}
  {\bibinfo  {journal} {J. Phys. B}\ }\textbf {\bibinfo {volume} {35}},\
  \bibinfo {pages} {3095} (\bibinfo {year} {2002})}\BibitemShut {NoStop}%
\bibitem [{\citenamefont {J{\"{o}}rdens}\ \emph {et~al.}(2008)\citenamefont
  {J{\"{o}}rdens}, \citenamefont {Strohmaier}, \citenamefont {G{\"{u}}nter},
  \citenamefont {Moritz},\ and\ \citenamefont {Esslinger}}]{Jordens2008}%
  \BibitemOpen
  \bibfield  {author} {\bibinfo {author} {\bibfnamefont {R.}~\bibnamefont
  {J{\"{o}}rdens}}, \bibinfo {author} {\bibfnamefont {N.}~\bibnamefont
  {Strohmaier}}, \bibinfo {author} {\bibfnamefont {K.}~\bibnamefont
  {G{\"{u}}nter}}, \bibinfo {author} {\bibfnamefont {H.}~\bibnamefont
  {Moritz}}, \ and\ \bibinfo {author} {\bibfnamefont {T.}~\bibnamefont
  {Esslinger}},\ }\href {\doibase 10.1038/nature07244} {\bibfield  {journal}
  {\bibinfo  {journal} {Nature}\ }\textbf {\bibinfo {volume} {455}},\ \bibinfo
  {pages} {204} (\bibinfo {year} {2008})}\BibitemShut {NoStop}%
\bibitem [{\citenamefont {Schulte}\ \emph {et~al.}(2005)\citenamefont
  {Schulte}, \citenamefont {Drenkelforth}, \citenamefont {Kruse}, \citenamefont
  {Ertmer}, \citenamefont {Arlt}, \citenamefont {Sacha}, \citenamefont
  {Zakrzewski},\ and\ \citenamefont {Lewenstein}}]{Schulte2005}%
  \BibitemOpen
  \bibfield  {author} {\bibinfo {author} {\bibfnamefont {T.}~\bibnamefont
  {Schulte}}, \bibinfo {author} {\bibfnamefont {S.}~\bibnamefont
  {Drenkelforth}}, \bibinfo {author} {\bibfnamefont {J.}~\bibnamefont {Kruse}},
  \bibinfo {author} {\bibfnamefont {W.}~\bibnamefont {Ertmer}}, \bibinfo
  {author} {\bibfnamefont {J.}~\bibnamefont {Arlt}}, \bibinfo {author}
  {\bibfnamefont {K.}~\bibnamefont {Sacha}}, \bibinfo {author} {\bibfnamefont
  {J.}~\bibnamefont {Zakrzewski}}, \ and\ \bibinfo {author} {\bibfnamefont
  {M.}~\bibnamefont {Lewenstein}},\ }\href {\doibase
  10.1103/PhysRevLett.95.170411} {\bibfield  {journal} {\bibinfo  {journal}
  {Phys. Rev. Lett.}\ }\textbf {\bibinfo {volume} {95}},\ \bibinfo {pages}
  {170411} (\bibinfo {year} {2005})}\BibitemShut {NoStop}%
\bibitem [{\citenamefont {White}\ \emph {et~al.}(2009)\citenamefont {White},
  \citenamefont {Pasienski}, \citenamefont {McKay}, \citenamefont {Zhou},
  \citenamefont {Ceperley},\ and\ \citenamefont {DeMarco}}]{White2009}%
  \BibitemOpen
  \bibfield  {author} {\bibinfo {author} {\bibfnamefont {M.}~\bibnamefont
  {White}}, \bibinfo {author} {\bibfnamefont {M.}~\bibnamefont {Pasienski}},
  \bibinfo {author} {\bibfnamefont {D.}~\bibnamefont {McKay}}, \bibinfo
  {author} {\bibfnamefont {S.~Q.}\ \bibnamefont {Zhou}}, \bibinfo {author}
  {\bibfnamefont {D.}~\bibnamefont {Ceperley}}, \ and\ \bibinfo {author}
  {\bibfnamefont {B.}~\bibnamefont {DeMarco}},\ }\href {\doibase
  10.1103/PhysRevLett.102.055301} {\bibfield  {journal} {\bibinfo  {journal}
  {Phys. Rev. Lett.}\ }\textbf {\bibinfo {volume} {102}},\ \bibinfo {pages}
  {055301} (\bibinfo {year} {2009})}\BibitemShut {NoStop}%
\bibitem [{\citenamefont {Choi}\ \emph {et~al.}(2016)\citenamefont {Choi},
  \citenamefont {Hild}, \citenamefont {Zeiher}, \citenamefont {Schauss},
  \citenamefont {Rubio-Abadal}, \citenamefont {Yefsah}, \citenamefont
  {Khemani}, \citenamefont {Huse}, \citenamefont {Bloch},\ and\ \citenamefont
  {Gross}}]{Choi2016}%
  \BibitemOpen
  \bibfield  {author} {\bibinfo {author} {\bibfnamefont {J.-Y.}\ \bibnamefont
  {Choi}}, \bibinfo {author} {\bibfnamefont {S.}~\bibnamefont {Hild}}, \bibinfo
  {author} {\bibfnamefont {J.}~\bibnamefont {Zeiher}}, \bibinfo {author}
  {\bibfnamefont {P.}~\bibnamefont {Schauss}}, \bibinfo {author} {\bibfnamefont
  {A.}~\bibnamefont {Rubio-Abadal}}, \bibinfo {author} {\bibfnamefont
  {T.}~\bibnamefont {Yefsah}}, \bibinfo {author} {\bibfnamefont
  {V.}~\bibnamefont {Khemani}}, \bibinfo {author} {\bibfnamefont {D.~A.}\
  \bibnamefont {Huse}}, \bibinfo {author} {\bibfnamefont {I.}~\bibnamefont
  {Bloch}}, \ and\ \bibinfo {author} {\bibfnamefont {C.}~\bibnamefont
  {Gross}},\ }\href {\doibase 10.1126/science.aaf8834} {\bibfield  {journal}
  {\bibinfo  {journal} {Science}\ }\textbf {\bibinfo {volume} {352}},\ \bibinfo
  {pages} {1547} (\bibinfo {year} {2016})}\BibitemShut {NoStop}%
\bibitem [{\citenamefont {Pezz{\`{e}}}\ \emph {et~al.}(2004)\citenamefont
  {Pezz{\`{e}}}, \citenamefont {Pitaevskii}, \citenamefont {Smerzi},
  \citenamefont {Stringari}, \citenamefont {Modugno}, \citenamefont
  {de~Mirandes}, \citenamefont {Ferlaino}, \citenamefont {Ott}, \citenamefont
  {Roati},\ and\ \citenamefont {Inguscio}}]{Pezze2004}%
  \BibitemOpen
  \bibfield  {author} {\bibinfo {author} {\bibfnamefont {L.}~\bibnamefont
  {Pezz{\`{e}}}}, \bibinfo {author} {\bibfnamefont {L.}~\bibnamefont
  {Pitaevskii}}, \bibinfo {author} {\bibfnamefont {A.}~\bibnamefont {Smerzi}},
  \bibinfo {author} {\bibfnamefont {S.}~\bibnamefont {Stringari}}, \bibinfo
  {author} {\bibfnamefont {G.}~\bibnamefont {Modugno}}, \bibinfo {author}
  {\bibfnamefont {E.}~\bibnamefont {de~Mirandes}}, \bibinfo {author}
  {\bibfnamefont {F.}~\bibnamefont {Ferlaino}}, \bibinfo {author}
  {\bibfnamefont {H.}~\bibnamefont {Ott}}, \bibinfo {author} {\bibfnamefont
  {G.}~\bibnamefont {Roati}}, \ and\ \bibinfo {author} {\bibfnamefont
  {M.}~\bibnamefont {Inguscio}},\ }\href {\doibase
  10.1103/PhysRevLett.93.120401} {\bibfield  {journal} {\bibinfo  {journal}
  {Phys. Rev. Lett.}\ }\textbf {\bibinfo {volume} {93}},\ \bibinfo {pages}
  {120401} (\bibinfo {year} {2004})}\BibitemShut {NoStop}%
\bibitem [{\citenamefont {Ott}\ \emph {et~al.}(2004)\citenamefont {Ott},
  \citenamefont {de~Mirandes}, \citenamefont {Ferlaino}, \citenamefont {Roati},
  \citenamefont {Modugno},\ and\ \citenamefont {Inguscio}}]{Ott2004}%
  \BibitemOpen
  \bibfield  {author} {\bibinfo {author} {\bibfnamefont {H.}~\bibnamefont
  {Ott}}, \bibinfo {author} {\bibfnamefont {E.}~\bibnamefont {de~Mirandes}},
  \bibinfo {author} {\bibfnamefont {F.}~\bibnamefont {Ferlaino}}, \bibinfo
  {author} {\bibfnamefont {G.}~\bibnamefont {Roati}}, \bibinfo {author}
  {\bibfnamefont {G.}~\bibnamefont {Modugno}}, \ and\ \bibinfo {author}
  {\bibfnamefont {M.}~\bibnamefont {Inguscio}},\ }\href {\doibase
  10.1103/PhysRevLett.92.160601} {\bibfield  {journal} {\bibinfo  {journal}
  {Phys. Rev. Lett.}\ }\textbf {\bibinfo {volume} {92}},\ \bibinfo {pages}
  {160601} (\bibinfo {year} {2004})}\BibitemShut {NoStop}%
\bibitem [{\citenamefont {Modugno}\ \emph {et~al.}(2003)\citenamefont
  {Modugno}, \citenamefont {Ferlaino}, \citenamefont {Heidemann}, \citenamefont
  {Roati},\ and\ \citenamefont {Inguscio}}]{Modugno2003}%
  \BibitemOpen
  \bibfield  {author} {\bibinfo {author} {\bibfnamefont {G.}~\bibnamefont
  {Modugno}}, \bibinfo {author} {\bibfnamefont {F.}~\bibnamefont {Ferlaino}},
  \bibinfo {author} {\bibfnamefont {R.}~\bibnamefont {Heidemann}}, \bibinfo
  {author} {\bibfnamefont {G.}~\bibnamefont {Roati}}, \ and\ \bibinfo {author}
  {\bibfnamefont {M.}~\bibnamefont {Inguscio}},\ }\href {\doibase
  10.1103/PhysRevA.68.011601} {\bibfield  {journal} {\bibinfo  {journal} {Phys.
  Rev. A}\ }\textbf {\bibinfo {volume} {68}},\ \bibinfo {pages} {011601(R)}
  (\bibinfo {year} {2003})}\BibitemShut {NoStop}%
\bibitem [{\citenamefont {Rigol}\ and\ \citenamefont
  {Muramatsu}(2004)}]{Rigol2004}%
  \BibitemOpen
  \bibfield  {author} {\bibinfo {author} {\bibfnamefont {M.}~\bibnamefont
  {Rigol}}\ and\ \bibinfo {author} {\bibfnamefont {A.}~\bibnamefont
  {Muramatsu}},\ }\href {\doibase 10.1103/PhysRevA.70.043627} {\bibfield
  {journal} {\bibinfo  {journal} {Phys. Rev. A}\ }\textbf {\bibinfo {volume}
  {70}},\ \bibinfo {pages} {043627} (\bibinfo {year} {2004})}\BibitemShut
  {NoStop}%
\bibitem [{\citenamefont {Ruuska}\ and\ \citenamefont
  {T{\"{o}}rm{\"{a}}}(2004)}]{Ruuska2004}%
  \BibitemOpen
  \bibfield  {author} {\bibinfo {author} {\bibfnamefont {V.}~\bibnamefont
  {Ruuska}}\ and\ \bibinfo {author} {\bibfnamefont {P.}~\bibnamefont
  {T{\"{o}}rm{\"{a}}}},\ }\href {\doibase 10.1088/1367-2630/6/1/059} {\bibfield
   {journal} {\bibinfo  {journal} {New J. Phys.}\ }\textbf {\bibinfo {volume}
  {6}},\ \bibinfo {pages} {59} (\bibinfo {year} {2004})}\BibitemShut {NoStop}%
\bibitem [{\citenamefont {Hooley}\ and\ \citenamefont
  {Quintanilla}(2004)}]{Hooley2004}%
  \BibitemOpen
  \bibfield  {author} {\bibinfo {author} {\bibfnamefont {C.~A.}\ \bibnamefont
  {Hooley}}\ and\ \bibinfo {author} {\bibfnamefont {J.}~\bibnamefont
  {Quintanilla}},\ }\href {\doibase 10.1103/PhysRevLett.93.080404} {\bibfield
  {journal} {\bibinfo  {journal} {Phys. Rev. Lett.}\ }\textbf {\bibinfo
  {volume} {93}},\ \bibinfo {pages} {080404} (\bibinfo {year}
  {2004})}\BibitemShut {NoStop}%
\bibitem [{Note1()}]{Note1}%
  \BibitemOpen
  \bibinfo {note} {In the cases where the exponentially small energy splitting
  between the symmetric and antisymmetric eigenstates of the problem is beyond
  the resolution of our numerical solver, we have `manually' taken linear
  combinations of the results to produce the correct symmetric and
  antisymmetric eigenstates}\BibitemShut {NoStop}%
\bibitem [{\citenamefont {Pezz{\`{e}}}\ \emph {et~al.}(2009)\citenamefont
  {Pezz{\`{e}}}, \citenamefont {Hambrecht},\ and\ \citenamefont
  {Sanchez-Palencia}}]{Pezze2009}%
  \BibitemOpen
  \bibfield  {author} {\bibinfo {author} {\bibfnamefont {L.}~\bibnamefont
  {Pezz{\`{e}}}}, \bibinfo {author} {\bibfnamefont {B.}~\bibnamefont
  {Hambrecht}}, \ and\ \bibinfo {author} {\bibfnamefont {L.}~\bibnamefont
  {Sanchez-Palencia}},\ }\href {\doibase 10.1209/0295-5075/88/30009} {\bibfield
   {journal} {\bibinfo  {journal} {EPL}\ }\textbf {\bibinfo {volume} {88}},\
  \bibinfo {pages} {30009} (\bibinfo {year} {2009})}\BibitemShut {NoStop}%
\bibitem [{\citenamefont {Gradshteyn}\ and\ \citenamefont
  {Ryzhik}(2000)}]{Gradshteyn}%
  \BibitemOpen
  \bibfield  {author} {\bibinfo {author} {\bibfnamefont {I.~S.}\ \bibnamefont
  {Gradshteyn}}\ and\ \bibinfo {author} {\bibfnamefont {I.~M.}\ \bibnamefont
  {Ryzhik}},\ }\href@noop {} {\emph {\bibinfo {title} {Table of Integrals,
  Series, and Products}}},\ \bibinfo {edition} {6th}\ ed.\ (\bibinfo
  {publisher} {Academic Press},\ \bibinfo {year} {2000})\ p.\ \bibinfo {pages}
  {797}\BibitemShut {NoStop}%
\bibitem [{\citenamefont {Kim}\ and\ \citenamefont {Zubarev}(2004)}]{Kim2004}%
  \BibitemOpen
  \bibfield  {author} {\bibinfo {author} {\bibfnamefont {Y.~E.}\ \bibnamefont
  {Kim}}\ and\ \bibinfo {author} {\bibfnamefont {A.~L.}\ \bibnamefont
  {Zubarev}},\ }\href {\doibase 10.1103/PhysRevA.70.033612} {\bibfield
  {journal} {\bibinfo  {journal} {Phys. Rev. A}\ }\textbf {\bibinfo {volume}
  {70}},\ \bibinfo {pages} {033612} (\bibinfo {year} {2004})}\BibitemShut
  {NoStop}%
\bibitem [{\citenamefont {Rigol}\ \emph {et~al.}(2006)\citenamefont {Rigol},
  \citenamefont {Muramatsu},\ and\ \citenamefont {Olshanii}}]{Rigol2006}%
  \BibitemOpen
  \bibfield  {author} {\bibinfo {author} {\bibfnamefont {M.}~\bibnamefont
  {Rigol}}, \bibinfo {author} {\bibfnamefont {A.}~\bibnamefont {Muramatsu}}, \
  and\ \bibinfo {author} {\bibfnamefont {M.}~\bibnamefont {Olshanii}},\ }\href
  {\doibase 10.1103/PhysRevA.74.053616} {\bibfield  {journal} {\bibinfo
  {journal} {Phys. Rev. A}\ }\textbf {\bibinfo {volume} {74}},\ \bibinfo
  {pages} {053616} (\bibinfo {year} {2006})}\BibitemShut {NoStop}%
\bibitem [{Note2()}]{Note2}%
  \BibitemOpen
  \bibinfo {note} {M. Schulz, F. Pollmann, C.A. Hooley, and R. Moessner, in
  preparation}\BibitemShut {NoStop}%
\bibitem [{\citenamefont {Peschel}(2012)}]{Peschel2012}%
  \BibitemOpen
  \bibfield  {author} {\bibinfo {author} {\bibfnamefont {I.}~\bibnamefont
  {Peschel}},\ }\href {\doibase 10.1007/s13538-012-0074-1} {\bibfield
  {journal} {\bibinfo  {journal} {Brazilian J. Phys.}\ }\textbf {\bibinfo
  {volume} {42}},\ \bibinfo {pages} {267} (\bibinfo {year} {2012})}\BibitemShut
  {NoStop}%
\bibitem [{\citenamefont {Blaizot}(1986)}]{Blaizot1986}%
  \BibitemOpen
  \bibfield  {author} {\bibinfo {author} {\bibfnamefont {J.}~\bibnamefont
  {Blaizot}},\ }\href@noop {} {\emph {\bibinfo {title} {Quantum Theory of
  Finite Systems}}}\ (\bibinfo  {publisher} {MIT Press},\ \bibinfo {year}
  {1986})\ p.~\bibinfo {pages} {38}\BibitemShut {NoStop}%
\end{thebibliography}%

\end{document}